\documentclass[12pt]{article}

\usepackage{amsfonts,ulem,url} \usepackage{amssymb} \usepackage{amsmath}
\usepackage{amstext,verbatim,graphicx,ifthen,multirow,amsthm}
\usepackage{subcaption} 
\usepackage{booktabs}
\usepackage{hyperref}
\usepackage{algorithm}
\usepackage{xcolor} 
\usepackage{natbib}
\usepackage{verbatim}
\usepackage[noend]{algpseudocode}

\usepackage{float}

\graphicspath{{figures/}} 

\renewcommand{\thepage}{}
\renewcommand{\appendix}{\footnotesize\parindent 0cm\setcounter{equation}{0} 
\renewcommand{\theequation}{A.\arabic{equation}}
\setcounter{lemma}{0}\renewcommand{\thelemma}{A.\arabic{lemma}}}

\newcommand{\bwexp}{\mathbf{W}^{\eexp}}

\newcommand{\bxexp}{\mathbf{X}^{\eexp}}

\newcommand{\bxexpn}{\mathbf{X}^{\eexp}_0}

\newcommand{\bxexpe}{\mathbf{X}^{\eexp}_1}

\newcommand{\bwexpn}{\mathbf{W}^{\eexp}_0}

\newcommand{\bwexpe}{\mathbf{W}^{\eexp}_1}

\newcommand{\byexpn}{\mathbf{Y}^{\eexp}_0}

\newcommand{\byexpe}{\mathbf{Y}^{\eexp}_1}

\newcommand{\cm}{{\rm cm}}

\newcommand{\dm}{{\rm dm}}
\newcommand{\dr}{{\rm dr}}

\newcommand{\eexp}{{\rm exp}}
\newcommand{\ecps}{{\rm cps}}
\newcommand{\epsid}{{\rm psid}}

\newcommand{\hoth}{{\rm ht}}

\newcommand{\lm}{{\rm lm}}

\newcommand{\nexp}{N^{\eexp}}
\newcommand{\nexpn}{N^{\eexp}_0}
\newcommand{\nexpe}{N^{\eexp}_1}
\newcommand{\ncps}{N^{\ecps}}
\newcommand{\ncpsn}{N^{\ecps}_0}

\newcommand{\npsid}{N^{\epsid}}
\newcommand{\npsidn}{N^{\epsid}_0}

\newcommand{\nnn}{{\rm nn}}

\newcommand{\mmx}{\mathbb{X}}
\newcommand{\mmz}{\mathbb{Z}}

\newcommand{\rf}{{\rm rf}}

\newcommand{\indep}{\perp\!\!\!\perp}

\newcommand{\mme}{\mathbb{E}}

\newcommand{\mmp}{\mathbb{P}}

\def\monthname{\ifcase\month\or
January\or February\or March\or April\or May\or June\or
July\or August\or September\or October\or November\or December\fi}

\renewcommand{\appendix}{\small\parindent 0cm\setcounter{equation}{0} 
\renewcommand{\theequation}{A.\arabic{equation}}
\setcounter{lemma}{0}\renewcommand{\thelemma}{A.\arabic{lemma}}
\setcounter{theorem}{0}\renewcommand{\thetheorem}{A.\arabic{theorem}}}

\begin{document}

\title{Using Wasserstein Generative Adversarial Networks for the Design of Monte Carlo Simulations\thanks{%
{\small 
We are grateful for comments by  participants in the conference in honor of Whitney Newey in April 2019, and especially for discussions with Whitney about econometrics over many years. 
Financial support from the Sloan Foundation and the Office of Naval Research under grant N00014-17-1-2131 is gratefully acknowledged.
We also want to acknowledge exceptional research assistance by Cole Kissane and Carolin Thomas. 
Code for estimation in this paper, as well as the the data used in the current paper, is available at \texttt{https://github.com/gsbDBI/ds-wgan}. 
}}}
\author{
Susan Athey\thanks{{\small Graduate School of Business, Stanford University, and NBER. Electronic correspondence: athey@stanford.edu. }}
\and
\and
 Guido W. Imbens\thanks{{\small Graduate School of Business, and Department of Economics,  Stanford University, and NBER. Electronic correspondence: imbens@stanford.edu.}} 
\and
Jonas Metzger\thanks{{\small Department of Economics, Stanford University. Electronic correspondence: metzgerj@stanford.edu. }} 
\and
Evan Munro\thanks{{\small  Graduate School of Business, Stanford University. Electronic correspondence: munro@stanford.edu,}} 
}
\date{
 \ifcase\month\or
January\or February\or March\or April\or May\or June\or
July\or August\or September\or October\or November\or December\fi \ \number%
\year
}
\maketitle

\begin{abstract}

When researchers develop new econometric methods it is common practice to compare the performance of the new methods to those of existing methods in Monte Carlo studies.
The credibility of such Monte Carlo studies is often limited because of the  discretion the researcher has in choosing the Monte Carlo designs reported. To improve the credibility we propose using a class of generative models that has recently been developed in the machine learning literature, termed Generative Adversarial Networks (GANs) which can be used to systematically generate artificial data that closely mimics existing datasets. Thus, in combination with existing real data sets, GANs can be used to limit the degrees of freedom in Monte Carlo study designs for the researcher, making any comparisons more convincing.
In addition, if an applied researcher is concerned with the performance of a particular statistical method on a specific data set (beyond its theoretical properties in large samples), she can use such GANs to assess the performance of the proposed method, {\it e.g.,} the coverage rate of confidence intervals or the bias of the estimator, using simulated data which closely resembles the exact setting of interest.
To illustrate these methods we apply Wasserstein GANs (WGANs) to the estimation of average treatment effects. In this example, we find that $(i)$ there is not a single estimator that outperforms the others in all three settings, so researchers should tailor their analytic approach to a given setting, $(ii)$ systematic simulation studies can be helpful for selecting among competing methods in this situation, and $(iii)$ the generated data closely resemble the actual data.
\end{abstract}


\baselineskip=20pt\newpage \setcounter{page}{1}\renewcommand{\thepage}{[%
\arabic{page}]}\renewcommand{\theequation}{\arabic{section}.%
\arabic{equation}}

\section{Introduction}
\label{section:introduction}

There has been rapid progress in the development of predictive statistical methods in recent years, particularly in the field of machine learning. This progress has been aided by the availability of a large number of benchmark real-world data sets. Specifying the criterion of out-of-sample predictive accuracy on these datasets defines a shared objective for the scientific community that new developments can be evaluated against. This does not work directly in  econometrics because the main objective of many econometric methods is the estimation of causal effects. Because the true causal effects are unobserved in real-world data sets, such data sets cannot directly serve as benchmarks to evaluate the performance of causal inference methods. 

Partly as a result, there is a long standing tradition in econometrics to compare the performance of the new methods to those of existing methods in Monte Carlo studies where researchers do have access to the true causal effect that generated the data. In such Monte Carlo studies, artificial data are frequently generated using  researcher-chosen distributions with a high degree of smoothness and limited dependence between different variables.  Because of the discretion the researcher has in chosen these distributions the performance of new methods in those settings is not always viewed as indicative of the performance in the real world. For  recent discussions of these issues, see \citet{advani2019mostly} and \citet{knaus2018machine}. 

In a similar but distinct setting, an applied researcher may have to decide on which particular statistical method to use on a specific data set. To make this decision, evidence on the properties of various estimators in a particular finite-sample setting can be useful, even when attractive theoretical properties in large sample are known to hold for some methods. In this situation, the researcher may wish to assess the performance, {\it e.g.,} the coverage rate of confidence intervals or the bias of an estimator using simulated data. For this purpose it would be helpful to able to  generate artificial data such that the distribution underlying  the simulations resembles the actual data set where the researcher wishes to implement the method.

In this paper we discuss how Generative Adversarial Nets (GANs, \citet{goodfellow2014generative}), and in particular GANs minimizing the Wasserstein distance (WGANs, \citet{arjovsky2017wasserstein}) can be used to systematically generate data that closely mimic real  data sets. Given an actual data set these methods allow researchers to systematically assess the performance of various estimators in  settings that are substantially more realistic than those often used in Monte Carlo studies. Moreover, by tying the data generating process to real data sets they can at least partly pre-empt concerns that researchers chose particular simulation designs to favor their proposed methods.
Additionally, the resulting data generating distributions can be shown to satisfy certain privacy guarantees with respect to the data they were trained on under some modifications, see  \citet{xie2018differentially}. This would allow the scientific community to benefit from otherwise inaccessible confidential data sources.

After a brief review of WGANs we apply them to a classic data set in the program evaluation literature, originally put together by \citet{lalonde}. 
We use the specific sample subsequently recovered by \citet{dehejiawahba, dehejia2002propensity} that is available online on Dehejia's website. We refer to this as the Lalonde-Dehejia-Wahba (LDW) data set. 
The LDW data set has some special features which make it a challenging setting for estimating average treatment effects under unconfoundedness. It is thus an attractive starting point for comparing some of the many estimators proposed for this problem. First, we demonstrate how WGANs can generate artificial data in this setting. Second, we discuss how similar the generated data are to the actual sample. Third, we assess the properties of a set of estimators for average treatment effects. Finally, we present approaches to evaluate various robustness properties of these results, such as robustness to sampling variation, size of the original data and WGAN hyperparameters.

We use three specific samples created from the LDW data to create a range of settings. First, in what we call the LDW-E (experimental sample),  we use the data from the original experiment. Second, LDW-CPS (observational sample) contains the experimental treated units and data on individuals from the CPS comparison sample. Third, in the LDW-PSID (observational sample) we use the experimental treated units and data from the PSID comparison sample. In our analysis we compare the performance of thirteen estimators for the average effect for the treated proposed in the literature to the baseline estimator equal to the difference in means by treatment status.  Some of the thirteen estimators are based on flexible estimators of the conditional outcome means, of the propensity score, or of both. These estimators are based on generalized linear models, random forests and neural nets, as well as balancing methods.

\section{Wasserstein Generative Adversarial Networks}
\label{section:wgan}

In this section we briefly review Generative Adversarial Networks (GANs), and in particular GANs based on the Wasserstein distance for an audience familiar with standard statistical methods such as maximum likelihood estimation.  GANs were first introduced in 
\citet{goodfellow2014generative}, and Wasserstein GANs (WGANs for short) were introduced in \citet{arjovsky2017wasserstein}. See
\citet{gui2020review} for a recent review. 
These methods have not received much attention in the econometrics literature yet, with the exception of \citet{manresa2019}.
The context we are interested in is as follows.
We have a sample of $N$ observations on $d_X$-component vectors, $X_1,\ldots,X_N$, drawn randomly from a distribution with cumulative distribution function $\mmp(\cdot)$, density $p_X(\cdot)$ and domain $\mmx$. We are interested in  drawing new observations that are similar to samples drawn from this distribution, but not necessarily identical. 

\subsection{Conventional Approaches to Nonparametric Estimation of Distributions}

A conventional approach in econometrics is to estimate the distribution $p_X(\cdot)$  using kernel density estimation (\citet{silverman2018density, hardle1990applied, tsybakov2008introduction}). Given a bandwidth $h$ and a kernel $K(\cdot)$, the standard kernel density estimator is
\[ \hat p_X(x)=\frac{1}{Nh^{d_X}}\sum_{i=1}^N K\left( \frac{X_i-x }{h}\right).\]
Conventional kernels include the Gaussian kernel and the Epanechnikov kernel. 
Bandwidth choices have been proposed to minimize squared-error loss. Standard kernel density estimators perform poorly in high-dimensional settings, and when the true data distribution has bounded support. In finite samples, kernel density estimation can be susceptible to oversmoothing. 

An alternative approach to generate new samples that are similar to an existing sample is the bootstrap (\citet{efron1982jackknife, efron1994introduction}), which estimates the cumulative distribution function as
\[ \hat \mmp(x)=\frac{1}{N}\sum_{i=1}^N \mathbf{1}_{X_i\leq x}.\]
The main disadvantage of this method and the reason this does not work for our purposes is that it cannot generate observations that differ from those seen in the original sample. As a consequence a large generated  sample would contain an unrealistic amount of identical data points. In the particular problem we study this would lead to the difficulty that in the population the propensity score, the probability of receiving the treatment conditional on the covariates, would always be equal to zero or one instead of strictly between zero and one as is required for the properties of many of the estimators proposed in this literature.

\subsection{Generative Adversarial Networks}
\label{section:gan}

Generative Adversarial Networks (GANs) are a recently developed alternative approach to generating data that are similar to a particular data set (\citet{goodfellow2014generative, arjovsky2017towards}).  GANs can be thought of as implicitly estimating the distribution, although they do not directly produce estimates of the density or distribution function at a particular point. Instead, they generate data from the estimated distribution. They do so by optimizing the parameters of a model for the distribution of the  data generating process (DGP) called the {\it generator}, which is trained in a type of mini-max game against an adversarial model called the {\it discriminator} that attempts to distinguish between the generated data and the real data. We introduce both pieces individually before we bring them together. GANs have recently become popular in the machine learning literature, but have not received much attention yet in the econometrics literature. An exception is \citet{manresa2019} who use WGANs for estimation of structural models. \citet{liang2018well} and \citet{singh2018nonparametric} derive theoretical properties of the implied distributions obtained from this class of algorithms. Important takeaways from their convergence rates are that GANs can learn distributions as well as the traditional methods in general, while not suffering as much from the curse of dimensionality whenever the data approximately lies on a lower-dimensional manifold.

\subsubsection{Generator}
In contrast to more conventional approaches to specifying distributions, the generator is defined as a non-stochastic push-forward mapping $g(\cdot;\theta_g): \mmz \to \mmx$, where $\theta_g$ denote the parameters of the generator and $\mmz$ is some latent space with dimension $d_g$, which often is, but need not be equal to $d_X$, the dimension of  $X$. 
For any distribution $p_Z(\cdot)$ over $\mmz$, this mapping implicitly defines a distribution $p_{\theta_g}(\cdot)$ (referred to as push-forward measure in measure theory) over $\mmx$ via
 \[ \tilde{X}=g(Z;\theta_g), \ Z\sim p_Z(\cdot) \implies \tilde{X} \sim p_{\theta_g}(\cdot) .\] $p_Z(\cdot)$ is chosen by the researcher to be  simple to draw from ({\it e.g.,} multivariate uniform or normal) and kept fixed throughout training.  
Given $g(\cdot)$ with a finite-dimensional $\theta_g$, this simply defines a parametric model for $X$.
As a result one could in principle estimate $\theta_g$ by maximum likelihood. The difficulty in doing so is that it can be difficult to evaluate the implied log likelihood function. Instead the GAN approach follows a different path.
Estimates of $\theta_g$ are  obtained by minimizing a notion of distance between the empirical distributions of samples from $p_{\theta_g}$ and the original data. Before describing the details of the optimization, we  introduce  a simple concrete example where both $\mmz$ and $\mmx$ are  scalars. Let $p_Z(\cdot)$ be the density of a standard normal distribution to illustrate some of the concepts: 
\[Z \sim{\cal N}(0,1).\] Our generator is a simple shift: 
 \[g(z; \theta_g)=z + \theta_g,\] 
 where $\theta_g$ is scalar as well. With this special choice for the generator, we can express the implied distribution of $\tilde{X}$ in closed form: 
 \[ \tilde{X}=g(Z; \theta_g), \ Z \sim{\cal N}(0,1) \implies \tilde{X} \sim{\cal N}(\theta_g, 1).\] 
 In practice the generator is usually parametrized in a much more flexible manner using a neural network. In that case, the researcher would not have access to a closed form expression of $p_{\theta_g}(\cdot)$, although typically it is still straightforward to draw samples from it given values for the parameter $\theta_g$. 

In this simple case estimating the parameter of the generator given the sample  is straighforward: the maximum likelihood estimator for $\theta_g$ is the sample average $\overline{X}$. However, we are interested in settings where 
$(i)$ the maximum likelihood estimator may be difficult to calculate, and
$(ii)$ the maximum likelihood estimator may not even be an attractive choice.

\subsubsection{Discriminator}
Next, we  explain how the GAN methodology estimates $\theta_g$ by minimizing a well defined notion of distance between the distribution of our model $p_{\theta_g}(\cdot)$ and that of the data $p_X(\cdot)$ without requiring a closed-form expression for either of the two. 
First, we examine  different notions of distances between distributions. For two distributions $\mmp$ and $\mmp'$ which are absolutely continuous with respect to the same measure $\mu(\cdot)$, the Kullback-Leibler divergence is given by
\[ KL(\mmp,\mmp')=\int \ln \left(\frac{\mmp(X)}{\mmp'(X)}\right) \mmp(X)d\mu(X).\] 
This distance is used with $\mmp$ equal to the empirical distribution in maximum likelihood estimation. In many settings this has attractive efficiency properties.
However, it has been argued that a  distance notion that is symmetric in $\mmp$ and $\mmp'$ would be preferable in the context of data generation (\citet{huszar2015not}): whereas maximum likelihood favors  values for the parameters under which the data from the empirical distribution is likely, data generation might be more interested in the reverse: values for the parameters which produce data that is likely under the empirical distribution. Intuitively, this can be related to the perceived tendency of KL-minimizers to {\it oversmooth} relative to the true distribution. An objective that addresses these concerns would be the Jensen-Shannon divergence \[ JS(\mmp,\mmp')=KL\left(\mmp\left| \frac{\mmp+\mmp'}{2}\right.\right)+KL\left(\mmp'\left| \frac{\mmp+\mmp'}{2}\right.\right).\]
 \citet{goodfellow2014generative} show that the JS divergence can be equivalently written as the solution to a particular optimization problem:
\[ JS(\mmp,\mmp') = \ln2 + \sup_{d: \mmx \mapsto (0,1)}\left\{ \frac{1}{2} \mme_{x\sim\mmp} \ln(d(x)) + \frac{1}{2} \mme_{x \sim\mmp'} \ln(1-d(x))\right\}.\]
\citet{goodfellow2014generative} call the function $d(\cdot)$ the {\it discriminator}. It has a simple interpretation: imagine a data generating process that with equal probability samples $x$ from either $\mmp$ or $\mmp'$. Then the above objective function  for $d(\cdot)$ corresponds to the maximum likelihood objective of a model which tries to classify which of the two distributions $x$ was sampled from, $\mmp$ or $\mmp'$. In this case the optimal discriminator is \[ d^*(x) = P(x\sim\mmp | x) = \dfrac{\mmp(x)}{\mmp(x) + \mmp'(x)}.\] 
Note that if the two distributions $\mmp$ and $\mmp'$ are equal, the discriminator is flat as a function of $x$.

Let us examine how this applies to the simple one-dimensional example from the previous subsection. Assume the original data was generated from a one-dimensional Gaussian with mean $\mu$ and unit variance, 
 \[X \sim {\cal N}(\mu, 1).\] 
 Then, we can plug in the Gaussian densities into the expression above to obtain the optimal discriminator $d^*$:
 \[d^*(x) 
 = \sigma(\mu^2-\theta_g^2 + 2(\mu-\theta_g)x) = \sigma(\theta_{0d}^* + \theta_{1d}^*x), \]
 where 
  $\sigma(x) = \exp(x)/(1+\exp(x))$ and 
 for $\theta_{0d}^* =\mu^2-\theta_g^2$ and $ \theta_{1d}^*=2(\mu-\theta_g)$.
 A key insight is that we do not require an analytical expression for either of the two densities to obtain the optimal discriminator. We can simply parametrize a model for the discriminator $d(x; \theta_d) = \sigma(\theta_{0d} + \theta_{1d}x)$ and optimize the  likelihood of correctly classifying random samples from the two distributions. The maximum likelihood estimator for $\theta_d$ will converge to the optimal $\theta_d^*$ as the sample size for the generated data and the actual data set increases. This discriminator estimates the JS divergence between the current generator and original data distribution and allows us to obtain gradients with which we can optimize the generator as described in the next section. Even if the true and generator distributions are very complex, we can  approximate the optimal discriminator by maximizing the empirical analogue of the maximum likelihood objective with any sufficiently flexible function approximator $d(\cdot; \theta_d)$ taking values in $(0, 1)$. This yields the original GAN formulation.

\subsection{Original GAN} 

Let $X_1,\ldots,X_{N_R}$ denote the original data as before and let $Z_1,\ldots, Z_{N_F}$ be a large number of samples from the researcher-chosen $p_{Z}(\cdot)$. \citet{goodfellow2014generative} propose to jointly optimize for the discriminator and generator via the saddle-point objective
\[ \min_{\theta_g} \max_{\theta_d}L(\theta_d,\theta_g),\]
where the objective function is
\[L(\theta_d,\theta_g)=\frac{1}{N_R}\sum_{i=1}^{N_R }\ln d(X_i;\theta_d)+ \frac{1}{N_F}\sum_{i=1}^{N_F}  \ln \left[1-  d\Bigl(g(Z_i;\theta_g);\theta_d\Bigr)\right].\]
Both the generator and discriminator are fully parametric models, though typically very flexible ones, {\it e.g.,} neural networks (typically so flexible that they require regularization to avoid overfitting, as discussed below). The  joint optimization is carried out  by switching back and forth between updating $\theta_d$ and $\theta_g$ in the respective directions implied by the gradient of the objective $L(\theta_d,\theta_g)$. This procedure can be interpreted as a two player mini-max game with alternating better-response dynamics. Using the arguments from the previous subsection, the authors discuss assumptions under which this process converges to the saddle-point in which the discriminator yields the JS divergence, which the generator minimizes by mimicking the original data $p_X(\cdot)$. Further implementation details, including those on regularization and the choice of tuning parameters such as batch size, are discussed in Subsection \ref{section:algorithm}.  Let us examine how this would play out in our simplified one-dimensional example with $X_i{\cal N}(\mu,1)$. Since both the original and discriminator distributions are Gaussian with constant variance, we are justified to restrict the search space for the discriminator to that of linear logistic regression functions as argued before. After aninitialization  $(\theta_g^{(0)},\theta_d^{(0)})$, we can optimize the saddle-point objective by iterating between the following two steps. Given values $(\theta^k_d,\theta^k_g)$ after $k$ steps of the algorithm we update the two parameters:
\begin{enumerate}
\item Update the discriminator parameter $\theta_d$ as
\[\theta^{k+1}_d=\arg \max_{\theta_d} L(\theta_d,\theta_g).\]
\item Update the generator parameter $\theta_g$ by taking a small step (small learning rate $\alpha$) along the derivative:
\[ \theta_g^{k+1}=\theta^k_g- \alpha\frac{\partial }{\partial \theta_g} L(\theta_d^{k+1},\theta_g).\]
\end{enumerate}
After optimizing the discriminator at each step, we get an estimate of the JS divergence and its gradient at the {\it current} value of $\theta_g$. We thus need to re-optimize the discriminator after every gradient update of $\theta_g$. Particularly when the discriminator is a neural network, a practical implementation would simply update $\theta_d$ for a few gradient steps only instead of solving its optimization until convergence. In our particular example, given a sufficiently large number of draws from $P_Z(\cdot)$, the process will converge to the JS minimizing value $\theta_g=\overline{X}$ implying a discriminator with $\theta_d = (\ln(N_R/(N_R+N_F)), 0)$ which cannot do better than guessing a constant probability of $N_R/(N_R+N_F)$ of the data being real.
Note that in the saddlepoint optimization we may completely optimize the parameters of the discriminator, but we should only take small steps for the generator to ensure convergence.

\subsection{Wasserstein GANs}
\label{section:earth}

In practice, optimization of the original  \citet{goodfellow2014generative} GAN objective has proven to be computationally challenging. Difficulties can arise when the discriminator becomes too proficient early on in detecting generated observations, becoming ``flat'' around the samples from the generator and thus failing to provide useful gradient information to the generator with which to improve that. This is related to the fact that the JS divergence is infinite between two distributions with non-identical support, no matter how close the distributions are in terms of moments or quantiles. See \citet{gulrajani2017improved, arjovsky2017towards} for details. An attractive alternative to the Jensen-Shannon divergence is the Earth-Mover or Wasserstein distance (\citet{arjovsky2017wasserstein}): \[ W(\mmp,\mmp')=\inf_{\gamma\in\Pi(\mmp,\mmp')}\mme_{(X,Y)\sim\gamma}\left[\|X-Y\|\right],\]
where $\Pi(\mmp,\mmp')$ is the set of joint distributions that have marginals equal to $\mmp$ and $\mmp'$. The term Earth-Mover distance comes from the interpretation that $W(\mmp,\mmp')$  is the amount of probability mass that needs to be transported  to move from the distribution $\mmp$ to the distribution $\mmp'$. The Earth-Mover/Wasserstein distance is symmetric and well-defined irrespective of the degree of overlap between the support of the distributions. \citet{arjovsky2017wasserstein} exploit the fact that the Wasserstein distance, like the JS divergence, admits a dual representation
\[ W(\mmp,\mmp')=\sup_{\|f\|_L\leq 1} \Bigl\{\mme_{X\sim \mmp}\left[ f(X)\right]-
\mme_{X\sim \mmp'}\left[ f(X)\right]\Bigr\},\]
where we take the supremum of the functions $f:\mathbb{X}\mapsto \mathbb{R}$ over all Lipschitz functions with Lipschitz constant equal to 1.
The function $f(\cdot)$ is known as the {\it critic} and its optimized value implies an upper bound on how much any Lipschitz-continuous moment can differ between the two distributions. We parametrize the critic as $f(x;\theta_c)$, using a flexible function form.
Ignoring the Lipschitz constraint, the empirical analogue of the optimization problem becomes
\begin{equation}\label{wgan} \min_{\theta_g}\max_{\theta_c}
\left\{\frac{1}{N_R}\sum_{i=1}^{N_R} f(X_i;\theta_c)-\frac{1}{N_F}\sum_{i=1}^{N_F} 
f(g(Z_i;\theta_g);\theta_c)
\right\}.\end{equation}
Given the generator, we choose the parameters of the critic  to maximize the difference between the average of $f(X_i;\theta_c)$ over the real data and the average over the generated data. We then choose the parameter of the generator $\theta_g$, to minimize this maximum difference. For this objective to be well-behaved, it is important to restrict the search to parameters that ensure that the critic is Lipschitz with constant 1. The original WGAN formulation considered parameter clipping to ensure this constraint, which causes computational problems. \citet{gulrajani2017improved} showed that these can be avoided by instead adding a penalty term to the objective function for the critic. This term directly penalizes the norm of the derivative of the critic $f(\cdot)$ with respect to its input along the lines connecting original and generated data points, which ensures the critic sufficiently satisfies the constraint. Specifically, the penalty term has the form
 \[ \lambda\left\{ \frac{1}{m}\sum_{i=1}^m\left[\max \left(0,  \left\|\nabla_{\hat x}  f\left(\hat{X}_{i};\theta_c\right)\right\|_{2}-1
\right)\right]^2\right\},\]
where the $\hat X_i=\epsilon_i X_i+(1-\epsilon_i)\tilde X_i$ are  random convex combinations of the real and  generated observations, with the $\varepsilon_i$ re-drawn at each step. Note that here we
do not use the full real data sample, but instead use random batches of the real and generated data of the same size $m$.

\subsection{The Algorithm}
\label{section:algorithm}

Instead of using all the data in each step of the algorithm, we repeatedly use random batches of the real data with batch size $m$,
denoted by $X_1,\ldots,X_m$, and each iteration generate the same number  $m$ of new fake observations from the input distribution, denoted by $Z_1,\ldots,Z_m$. 
The general algorithm is described in Algorithm \ref{algo:wgan}.
For the optimization we use a modification of the SGD (Stochastic Gradient Descent) algorithm ({\it e.g.,} \citet{bottou2010large}),  the Adam (Adaptive moment estimation, \citet{kingma2014adam}) algorithm. The Adam algorithm combines the estimate of the (stochastic) gradient with previous estimates of the gradient, and  scales this using an estimate of the second moment of the unit-level gradients. The latter part is somewhat akin to the way the Berndt-Hall-Hall-Hausman algorithm proposed in \citet{berndt1974estimation} rescales the first derivatives using the inverse of the outer product matrix of the observation-level gradients, with the difference that Adam only uses the inverses of the diagonal elements of the outer product matrix of the gradients. Details are provided in the appendix.
Our specific implementation uses dropout (\citet{warde2013empirical, wager2013dropout}) to regularize the generator, which sets a random sample of $q\%$ of the weights in the generator network to zero at each step of the training. Without regularization, the generator may get close to the empirical distribution function especially if the batch size is large.

\begin{algorithm}[htbp]
\caption{WGAN}\label{algo:wgan}
\begin{algorithmic}[1]
\\ {$\rhd$ Tuning parameters: }
\\ \hskip0.6cm {$m$, batch size}
\\ \hskip0.6cm {$n_{critic}=15$, number of critic iterations per iteration of the generator}
\\ \hskip0.6cm {$lr_0=0.0001$, $\beta_1=0.9,\beta_2=0.999,\epsilon=10^{-8}$, parameters for Adam algorithm with hypergradient descent} 
\\ \hskip0.6cm {$\lambda=5$, penalty parameter for derivative of critic}
\\ {$\rhd$ Starting Values: }
\\ \hskip0.6cm {$\theta_c=0$ (critic), $\theta_g=0$ (generator)}
\\ {$\rhd$ Noise Distribution: }
\\ \hskip0.6cm {$p_Z(z)$ is mean zero Gaussian with identity covariance matrix, dimension equal to that of $x$}
\\
\While{$\theta_g$ has not converged}\\
\hskip0.6cm  $\rhd$ {Run $n_{critic}$ training steps for the critic.}
\For{$t=0,...,n_{critic}$}
\State Sample $\left\{X_{i}\right\}^{m}_{i=1} \sim \mathcal{D}$ (a batch of size $m$ from the real data, without replacement)
\State Sample $\{Z_{i}\}^{m}_{i=1} \sim p_Z(z)$ noise.
\\
\hskip1.2cm  $\rhd$ {Generate $m$ fake observations from the noise observations. }
\State $\tilde{X}_i \gets g(Z_{i};\theta_g)$ for  $i=1,\ldots,m$
\\
\hskip1.2cm  $\rhd$ {Compute penalty term $Q(\theta_c) $. }
\State Generate $\epsilon_i$, $i=1,\ldots,m$ from uniform distribution on $[0,1]$
\State  Calculate $
\hat X_i=\epsilon_i X_i+(1-\epsilon_i) \tilde X_i$ convex combinations of real and fake observations
\State $Q(\theta_c) \gets \frac{1}{m}\sum_{i=1}^m \left[\max\left(0, \left\| \nabla_{\hat x} f\left(\hat{X}_{i};\theta_c\right)\right\|_2-1
\right)\right]^2$
\\
\hskip1.2cm  $\rhd$ {Compute gradient with respect to the critic parameter $\theta_c$. }
\State $\delta_{\theta_c} \gets \nabla_{\theta_c} \left\lbrack \frac{1}{m}\sum_{i=1}^m f\left(X_{i};\theta_c\right)
-\frac{1}{m}\sum_{i=1}^m f\left(\tilde{X}_{i};\theta_c\right)+
\lambda Q(\theta_c)\right\rbrack
$ \label{line:critic_loss}
\State $\theta_c \gets {\rm Adam}(-\delta_{\theta_c}, \theta_c,\alpha,\beta_1,\beta_2)$
(update critic parameter using Adam algorithm)
\State {\bf end for}
\EndFor
\\
\hskip0.6cm  $\rhd$ {Run a single generator training step.}

\State Sample $\{Z_{i}\}^{m}_{i=1} \sim p_Z(z)$ noise.
\\
\hskip0.6cm  $\rhd$ {Compute gradients with respect to the generator parameters. }
\State $\delta_{\theta_g} \gets \nabla_{\theta_g}\frac{1}{m}\sum_{i=1}^m f\left(
g(Z_{i};\theta_g);\theta_c\right)$
\State  $\theta_g \gets{\rm Adam}(\delta_{\theta_g},\theta_g,\alpha,\beta_1,\beta_2)$
 (update generator parameter using Adam algorithm)
 \State {\bf end while}
\EndWhile
\end{algorithmic}
\end{algorithm}

\subsection{Conditional WGANs}
\label{section:conditionalwgan}

The algorithm discussed  in Section \ref{section:algorithm} learns to generate draws from an unconditional distribution. In many cases we want to generate data from a conditional distribution. For example, for the causal settings that motivate this paper, we may wish to keep fixed the number of treated and control units. This would be simple to implement by training two unconditional WGANs. More importantly, we wish to generate potential treated and control outcomes given a common set of pre-treatment variables. For that reason it is important to generate data from a conditional distribution (\citet{mirza2014conditional, odena2017conditional, liu2018auto, kocaoglu2017causalgan}).

Suppose we have a sample of real data $(X_i,V_i)$, $i=1,\ldots,N_R$. We wish to train a generator to sample from the conditional distribution of $X_i|V_i$. The conditioning variables $V_i$ are often referred to as {\it labels} in this literature. This can be achieved under minimal modifications to the unconditional WGAN algorithm described before: we simply feed $V_i$ as input to both the generator and the discriminator/critic, by concatenating it to their respective input vectors (i.e. the noise $Z_{i}$ and the observations $X_i$ respectively). To illustrate why this works, let the conditioning variables take values in some finite set $V_i \in \mathbb{V}$ and apply the law of iterated expectations to the infinite-sample version of the GAN objective:

\begin{multline*}
\inf_g \sup_{\|f\|_L\leq 1} \Bigl\{\mme_{X, V}\left[ f(X, V)\right]-
\mme_{Z, V}\left[ f(g(Z, V), V)\right]  \Bigr\}= \\
 \sum_{v\in \mathbb{V}} P(V=v) \inf_{g_v} \sup_{\|{f_v}\|_L\leq 1} \Bigl\{ \mme_{X|V=v}\left[ f_v(X)\right]-
\mme_{Z|V=v}\left[ f_v(g_v(Z))\right] \Bigr\}
\end{multline*}

As long as we enforce the Lipschitz constraint on the critic only with respect to $X_i$ and do not otherwise restrict the functional forms of the discriminator and critic, the resulting infinite-sample objective therefore simply corresponds to fitting an independent WGAN for every value $v\in \mathbb{V}$. Of course, with finite samples, particularly in the continuous case, we will allow the models to benefit from the smoothness of the conditional distribution $X_i|V_i$ by limiting the flexibility of the parametric models for the generator $g(Z_i|V_i;\theta_g)$ and the critic $f(X_i|V_i;\theta_c)$. In this case the equivalence disappears, since the optimization is not performed separately for different values of $V_i\in\mathbb{V}$, but the intuition is similar. The specific algorithm is described in Algorithm \ref{algo:cwgan}.

\begin{algorithm}[htbp]
\caption{CWGAN}\label{algo:cwgan}
\begin{algorithmic}[1]
\\ {$\rhd$ Tuning parameters: }
\\ \hskip0.6cm {$m$, batch size}
\\ \hskip0.6cm {$n_{critic}=15$, number of critic iterations per iteration of the generator}
\\ \hskip0.6cm {$lr_0=0.0001$, $\beta_1=0.9,\beta_2=0.999,\epsilon=10^{-8}$, parameters for Adam algorithm with hypergradient descent} 
\\ \hskip0.6cm {$\lambda=5$, penalty parameter for derivative of critic}
\\
\\ {$\rhd$ Starting Values: }
\\ \hskip0.6cm {$\theta_c=0$ (critic), $\theta_g=0$ (generator)}
\\ {$\rhd$ Noise Distribution: }
\\ \hskip0.6cm {$p_Z(z)$ is mean zero Gaussian with identity covariance matrix, dimension equal to that of $x$}
\\
\While{$\theta$ has not converged}\\
\hskip0.6cm  $\rhd$ {Run $n_{critic}$ training steps for the critic.}
\For{$t=0,...,n_{critic}$}
\State Sample $\left\{\left(X_{i},V_{i}\right)\right\}^{m}_{i=1} \sim \mathcal{D}$ a batch from the real data and labels.
\State Sample $\{Z_i\}^{m}_{i=1} \sim p_Z(z)$ noise.
\\
\hskip1.2cm  $\rhd$ {Generate $m$ fake observations $\tilde X_i$ corresponding to the $m$ real labels $V_i$. }
\State $\tilde{X}_i \gets g(Z_{i}|V_{i};\theta_g)$ for each $i$
\\

\hskip1.2cm  $\rhd$ {Compute penalty term $Q(\theta_c) $. }
\State Generate $\epsilon_i$, $i=1,\ldots,m$ from uniform distribution on $[0,1]$
\State  Calculate $
\hat X_i=\epsilon_i X_i+(1-\epsilon_i) \tilde X_i$ convex combinations of real and fake observations

\State $Q(\theta_c) \gets \frac{1}{m}\sum_{i=1}^m \left[\max\left(0, \left\| \nabla_{\hat x} f\left(\tilde{X}_{i}|V_i;\theta_c\right)\right\|_2-1
\right)\right]^2$
\\
\hskip1.2cm  $\rhd$ {Compute gradient with respect to the critic parameter $\theta_c$. }
\State $\delta_{\theta_c} \gets \nabla_{\theta_c} \left\lbrack \frac{1}{m}\sum_{i=1}^m f\left(X_{i}|V_i;\theta_c\right)
-\frac{1}{m}\sum_{i=1}^m f\left(\tilde{X}_{i}|V_i;\theta_c\right)+
\lambda Q(\theta_c)\right\rbrack
$ \label{line:c_critic_loss}
\State $\theta_c \gets {\rm Adam}(-\delta_{\theta_c}, \theta_c,\alpha,\beta_1,\beta_2)$
(update critic parameter using Adam algorithm)
\State {\bf end for}
\EndFor
\\
\hskip0.6cm  $\rhd$  {Run a single generator training step.}
\State Sample $\{V_{i}\}^{m}_{i=1} \sim \mathcal{D}$ a batch of size $m$ from the real labels.
\State Sample $\{Z_{i}\}^{m}_{i=1} \sim p_Z(z)$ noise.
\\
\hskip0.6cm  $\rhd$  {Compute gradients with respect to the generator parameters. }
\State $\delta_{\theta_g} \gets \nabla_{\theta_g}
\frac{1}{m}\sum_{i=1}^m f\left(g(Z_i|V_i;\theta_g)|V_{i};\theta_c\right)$
\State  $\theta_g \gets{\rm Adam}(\delta_{\theta_g},\theta_g,\alpha,\beta_1,\beta_2)$
 (update generator parameter)
 \EndWhile
\end{algorithmic}
\end{algorithm}

\section{Simulating the Lalonde-Dehejia-Wahba Data}
\label{section:ldw}

In this section we discuss the applicaction of  WGANs for Monte Carlo studies based on the Lalonde-Dehejia-Wahba (LDW) data.

\subsection{Simulation Studies for Average Treatment Effects}

In the setting of interest we have data on an outcome $Y_i$, a set of pretreatment variables $X_i$ and a binary treatment $W_i\in\{0,1\}$. We postulate that there exists for each unit in the population two potential outcomes $Y_i(0)$ and $Y_i(1)$, with the observed outcome equal to corresponding to the potential outcome for the treatment received, $Y_i=Y_i(W_i)$. We are interested in the average treatment effect for the treated,
\[ \tau=\mme[Y_i(1)-Y_i(0)|W_i=1],\]
assuming unconfoundedness (\citet{rosenbaum1983central, imbens2015causal}):
\[ W_i\ \indep\ \Bigl(Y_i(0),Y_i(1)\Bigr)\ \Big|\ X_i,\]
and overlap
\[ 0<{\rm pr}(W_i=1|X_i=x)<1,\ \ \forall\ x,\]
in combination referred to as ignorability.
Let $\mu(w,x)\equiv\mme[Y_i|W_i=w,X_i=x] $ (which by unconfoundedness is equal to $\mme[Y_i(w)|X_i=x]$) be the conditional outcome mean, and let $e(x)\equiv {\rm pr}(W_i=1|X_i=x)$ be the propensity score. There is a large literature developing methods for estimating  average and conditional average treatment effects in this setting (see \citet{imbens2004, abadie2018econometric} for surveys).

In this setting, researchers have often conducted  simulation studies to assess the properties of proposed methods (\citet{athey2018approximate, belloni2014inference,   huber2013performance, lechner2013sensitivity, lechner2019practical,  wendling2018comparing}).
Most closely related in the spirit of creating  simulation designs that  closely resemble real data are \citet{abadie2011bias, schuler2017synth, knaus2018machine}.
Using the LDW sample
\citet{abadie2011bias} estimate a model for the conditional means and the propensity score allowing for linear terms and second order terms. To account for the mass points at zero, they model separately the probability of the outcome being equal to zero and outcome conditional on being positive.
\citet{schuler2017synth} also start with a real data set. They postulate a value for the conditional average treatment effect $\tau(x)=\mme[Y_i(1)-Y_i(0)|X_i=x]=\mu(1,x)-\mu(0,x)$. They then use the empirical distribution of $(W_i,X_i)$ as the true distribution. They estimate the conditional means $\mu(w,x)$ using 
flexible models, imposing the constraint implied by the choice of conditional average treatment effect $\tau(x)$.
Given these estimates they estimate the residual distribution as the empirical distribution of $Y_i-\hat\mu(W_i,X_i)$. Then they impute outcomes for new samples using the estimated regression functions and random draws from the empirical residual distribution. Note that this procedure imposes homoskedasticity.
Note also that the 
\citet{schuler2017synth}   choice for the joint distribution of $(W_i,X_i)$ can create violations of the overlap requirement if the pre-treatment variables $X_i$ are continuous. Because they specify the conditional average treatment effect that does not create problems for estimating the ground truth. 
\citet{knaus2018machine} develop what they call empirical Monte Carlo methods where they use the empirical distribution of the covariates and the control outcome, combined with postulated individual level treatment effects and a flexibly estimated propensity score to generate artificial data.

\subsection{The LDW Data}
\label{section:ldwdata}

The data set we use in this paper was originally
constructed by
\citet{lalonde}, and later recovered by \citet{dehejiawahba} and available on Dehejia's website.  This data set has been widely used in the program evaluation literature to compare different methods for estimating average treatment effects ({\it e.g.,} 
\citet{dehejia2002propensity, heckmanhotz, abadie2011bias, ma2010robust} and many others). 
We use three versions of the  data. The first, which we refer to as the experimental sample, LDW-E, contains the observations from the actual experiment. This sample contains $\nexp$ observations, with $\nexpn=260$ control observations and $\nexpe=\nexp-\nexpn=185$ treated observations. For each individual in this sample we observe a set of eight pre-treatment variables, denoted by $X_i$. These include two earnings measures, two indicators for ethnicity, marital status, and two education measures, and age.
$\bxexpn$ denotes the $\nexpn\times 8$ matrix with each row corresponding to the pre-treatment variables for one of these units, and  $\bxexpe$ denoting the $\nexpe\times 8$ for the treated units in this sample. Let $\bxexp$ denote the $\nexp\times 8$ matrix with all the covariates.  Simiarly, let $\byexpn$ denote the $\nexpn$ vector of outcomes for the control units in this sample, and  $\byexpe$ denote the $\nexpe$ vector of outcomes for the treated units, and let    $\bwexpn$ denote the $\nexpn$ vector of treatment indicators for the control units in this sample (all zeros), and  $\bwexpe$ denote the $\nexpe$ vector of outcomes for the treated units (all ones).
The outcome is a measure of earnings in 1978.

The second sample is the CPS sample, LDW-CPS. It combines the treated observations from the experimental sample with $\ncpsn=15,992$ control observations drawn from the Current Population Survey, for a total of $\ncps=\nexpe+\ncpsn=16,177$ observations.
The third sample is the PSID sample, LDW-PSID.  It combines the treated observations from the experimental sample with $\npsidn=2,490$ control observations drawn from the Panel Survey of Income Dynamics, for a total of $\npsid=\nexpe+\npsidn=2,675$ observations. 
Table \ref{tabel1} presents summary statistics for the eight pretreatment variables and the outcome  by treatment status  in these samples.

\begin{table}[ht]
 \caption{\sc  Summary Statistics   for Lalonde-Dehejia-Wahba Data}
 \vskip1cm
 \begin{center}
 \begin{tabular}{lcccccccc}
\hline\hline \\
& \multicolumn{2}{c}{Experimental } 
& \multicolumn{2}{c}{Experimental}
& \multicolumn{2}{c}{CPS}
& \multicolumn{2}{c}{PSID}\\
& \multicolumn{2}{c}{trainees (185)} 
&\multicolumn{2}{c}{controls  (260)} 
& \multicolumn{2}{c}{controls (15,992)}
& \multicolumn{2}{c}{controls (2,490)} \\
& mean & s.d. & mean & s.d.& mean & s.d. & mean & s.d.\\
\\ \hline \\
    {\tt black} &             0.84 &         (0.36) &              0.83 &          (0.38) &              0.07 &          (0.26) &               0.25 &           (0.43) \\
  {\tt hispanic} &             0.06 &         (0.24) &              0.11 &          (0.31) &              0.07 &          (0.26) &               0.03 &           (0.18) \\
       {\tt age} &            25.82 &         (7.16) &             25.05 &          (7.06) &             33.23 &         (11.05) &              34.85 &          (10.44) \\
   {\tt married} &             0.19 &         (0.39) &              0.15 &          (0.36) &              0.71 &          (0.45) &               0.87 &           (0.34) \\
  {\tt nodegree} &             0.71 &         (0.46) &              0.83 &          (0.37) &               0.3 &          (0.46) &               0.31 &           (0.46) \\
 {\tt education} &            10.35 &         (2.01) &             10.09 &          (1.61) &             12.03 &          (2.87) &              12.12 &           (3.08) \\
  {\tt earn '74} &              2.1 &         (4.89) &              2.11 &          (5.69) &             14.02 &          (9.57) &              19.43 &          (13.41) \\
  {\tt earn '75} &             1.53 &         (3.22) &              1.27 &           (3.1) &             13.65 &          (9.27) &              19.06 &           (13.6) \\
  {\tt earn '78} &             6.35 &         (7.87) &              4.55 &          (5.48) &             14.85 &          (9.65) &              21.55 &          (15.56) \\
&&\\ \hline
 \end{tabular}
 \end{center}
 \label{tabel1}
\end{table}

\subsection{A Conditional WGAN for the LDW Data}
\label{section:wgan_ldw}

Consider the experimental data set LDW-E. The goal is to create samples of $\nexp$ observations, containing $\nexpn=260$ control units and $\nexpe=185$ treated units, where the samples are similar to the real sample. We proceed as follows. First, we run a conditional WGAN on the sample $\bxexp$, conditional on $\bwexp$. Let the parameters of the generator of the WGAN be $\theta_{g,X}^{\eexp}$. During training of the models, each batch of training data contains the same fraction of treated to avoid estimation issues when the fraction treated is close to zero (for example, this fraction is equal to 0.011 in the CPS dataset). 

In each case, for the generator we use a neural net with the following architecture. There are three hidden layers in the neural net, with the number of inputs and outputs equal to $(d_X+M,128), $ $(128,128)$ and $(128,128)$ respectively. Here $d_X$ is the dimension of the  vectors whose distribution we are modeling and $M$ is the dimension of the conditioning variables. For generating the covariates  conditional on the treatment, this is $d_X=8$, and $M=1$, and for generating the outcome variable conditional on the treatment and covariates this is $d_X=1$, and $M=9$. We use the rectified linear transformation, $a(z)=z\mathbf{1}_{z>0}$ in the hidden layers.
For the final layer we have 128 inputs and $d_X$ outputs. Here we use for binary variables a sigmoid transformation, for censored variables a rectified linear transformation, and for continuous variables the identity function.
We experimented a bit with shallower neural nets, including single layer networks that are known to be able to approximate functions arbitrarily closely (\citet{chen1998sieve, chen1999improved}).
In this case with relatively modest sample sizes one would expect that the single layer networks 
would be competitive with the deeper networks, and this is consistent with our experience. However, the deeper architectures were less sensitive to the hyperparameter choices, which is an advantage even with our modest sample sizes. The current state of the literature suggests that in complex big data settings deeper networks outperform the shallower ones
(\citet{goodfellow2016deep, choromanska2015loss}).
 
For the critic we use the same architecture with three layers, with 
the number of inputs and outputs equal to
$(d_X+M,128), $ $(128,128)$ and $(128,128)$ respectively. 
For the final layer we have 128 inputs, and 1 linear output.

We use batch sizes of  128, 4096, and 512 for the experimental sample, the cps sample, and the psid  sample.

We did not adapt the architectures to the individual settings, so these hyperparameters should not be thought of as optimal. In spite of this, they yield a well-performing WGAN. This is to emphasize that the exact architectural choices do not matter in settings like ours, so long as the overall size of the network is large enough to capture the complexity of the data and the amount of regularization ({\it i.e.}, dropout probability) is high enough to avoid over-fitting. 

 Given the parameters for the generators, $\theta_{g,X|W}^{\eexp}$, $\theta_{g,Y(W)|X,W}^{\eexp}$, we first create a single very large sample, with $N=10^6$ units. We use this sample as our population for the simulations. To create the large sample, first we draw separately the covariates for the treated and control units using the generator with parameter $\theta_{g,X|W}^{\eexp}$. In this step, we create the sample keeping the fraction of treated units equal to that in the sample. Next we draw independently  $Y(0)$ and $Y(1)$ for each observation in this large sample, using the $X$ and $W$ as the conditioning variables, using the generators with parameters $\theta_{g,Y(W)|X,W}^{\eexp}$. Unlike in any real dataset, we observe both $Y(0)$ and $Y(1)$ for each unit, simplifying the task of estimating the ground truth in the simulated data.  We use this single large sample to calculate the approximate true average effect for the treated as the average difference between the two potential outcomes for the treated units:
\[ \tau=\frac{1}{N_1}\sum_{i:W_i=1} \Bigl(Y_i(1)-Y_i(0)\Bigr).\]
For this fixed population we report in Table \ref{tabel2} the means and standard deviations for the same ten variables as in Table \ref{tabel1}. The means and standard deviations are fairly similar.
However, 
the fact that the first two moments of the generated data closely match those of the actual data is only limited comfort.
There are simple ways in which to generate data for which the first two moments of each of the variables match exactly those of the actual data, such as the standard bootstrap or a multivariate normal distribution. However, our generator allows us to generate new samples that contain observations not seen in the actual data, and with no duplicate observations.  The latter is important in our setting as discussed before.
\begin{table}[t]
\caption{\label{tab:} \sc Summary Statistics for WGAN-Generated Data Based on LDW Data}
\vskip1cm 
\centering
\begin{tabular}{lcccccccccccc}
\hline
\hline 
\\ 
& \multicolumn{2}{c}{Experimental } 
& \multicolumn{2}{c}{Experimental}
& \multicolumn{2}{c}{CPS}
& \multicolumn{2}{c}{PSID}\\
& \multicolumn{2}{c}{trainees} 
&\multicolumn{2}{c}{controls} 
& \multicolumn{2}{c}{controls}
& \multicolumn{2}{c}{controls} \\
& mean & s.d. & mean & s.d.& mean & s.d. & mean & s.d.\\
\\ \hline\\
    {\tt black} &      0.92 &  (0.27) &      0.91 &  (0.29) &      0.08 &   (0.27) &       0.26 &   (0.44) \\
  {\tt hispanic} &      0.03 &  (0.18) &      0.03 &  (0.18) &      0.07 &   (0.25) &       0.03 &   (0.17) \\
       {\tt age} &     25.61 &  (6.44) &     25.54 &  (7.28) &     32.82 &  (10.98) &      34.95 &   (10.7) \\
   {\tt married} &      0.15 &  (0.36) &       0.1 &   (0.3) &      0.72 &   (0.45) &       0.87 &   (0.34) \\
  {\tt nodegree} &      0.69 &  (0.46) &      0.81 &  (0.39) &      0.29 &   (0.45) &        0.3 &   (0.46) \\
 {\tt education} &     10.35 &   (1.7) &     10.02 &  (1.62) &     12.03 &   (2.85) &      12.32 &   (3.11) \\
  {\tt earn '74} &      3.07 &  (6.73) &      2.13 &  (5.51) &     14.07 &    (9.6) &      19.86 &  (13.06) \\
  {\tt earn '75} &      2.06 &  (3.41) &      1.29 &  (2.69) &     13.73 &   (9.38) &      20.41 &  (13.21) \\
  {\tt earn '78} &      5.31 &  (5.49) &      3.97 &  (4.46) &     14.89 &   (9.82) &      22.62 &  (14.66) \\
\\ \hline
\end{tabular}
\label{tabel2}
\end{table}
In Figures 1-3 we present some graphical evidence on the comparison of the actual data and the generate data for the CPS control sample.
In general the generated data and the actual data are quite similar. This is true for not just for the first two moments, but also for the  marginal distributions, the correlations, as well as the conditional distributions. In particular it is impressive to see in Figure 5 the conditional distribution of 1978 earnings for two groups for the actual data (those with 1974 earnings positive or zero). These two conditional distributions of 1978 earnings have quite different shapes, yet both are well matched  in the two samples. A multivariate normal distribution could not have reproduced such patterns.

\begin{figure}[!ht]
\centering
\caption{\sc Marginal Histograms for CPS Data}
\begin{subfigure} [] {0.29\textwidth} 
\includegraphics[width=\textwidth]{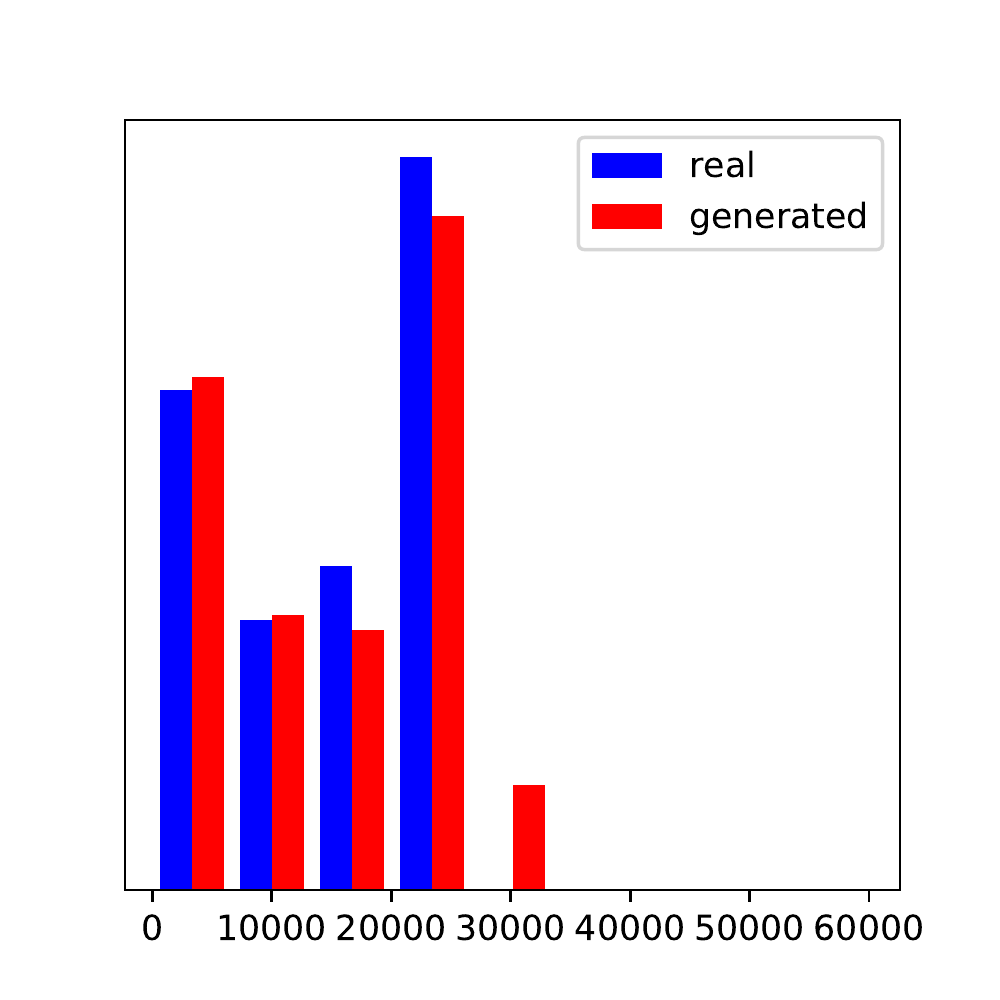}
\caption{Earnings 1978}
\end{subfigure} 
\begin{subfigure} [] {0.29\textwidth} 
\includegraphics[width=\textwidth]{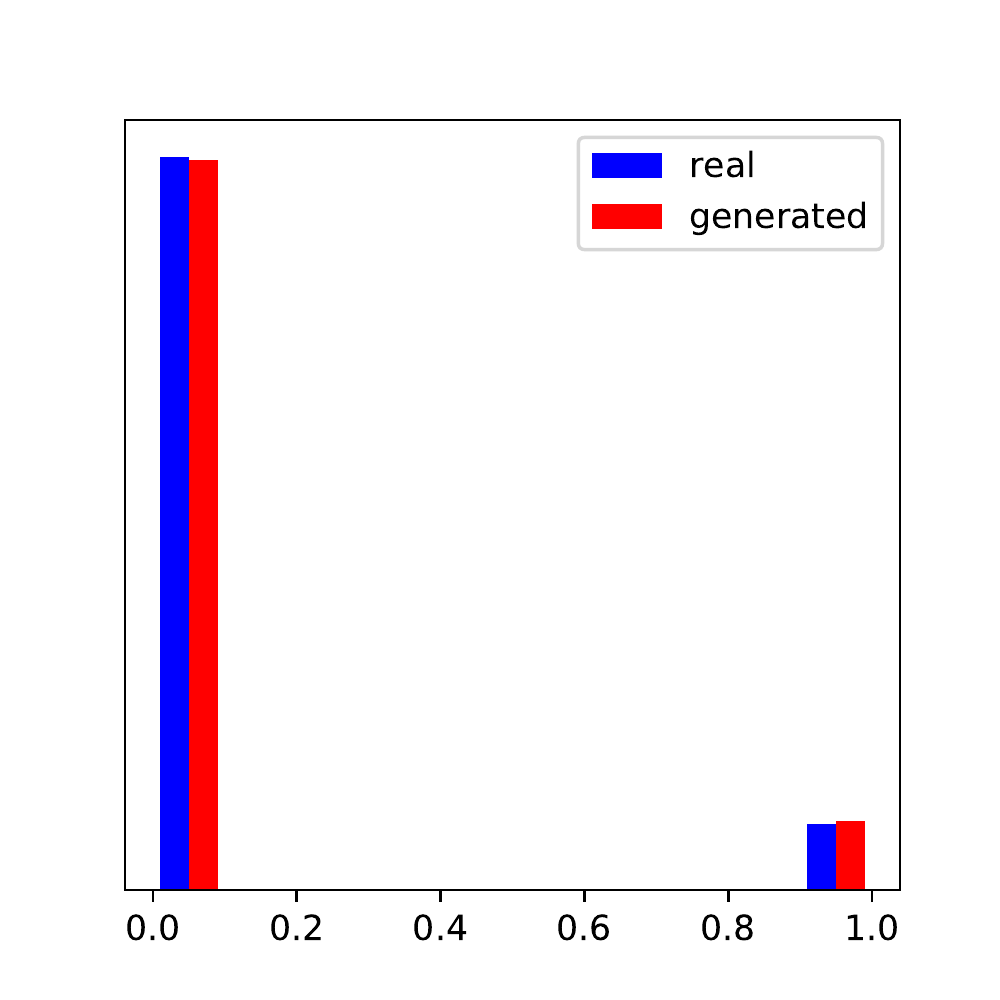}
\caption{Black}
\end{subfigure} 
\begin{subfigure} [] {0.29\textwidth} 
\includegraphics[width=\textwidth]{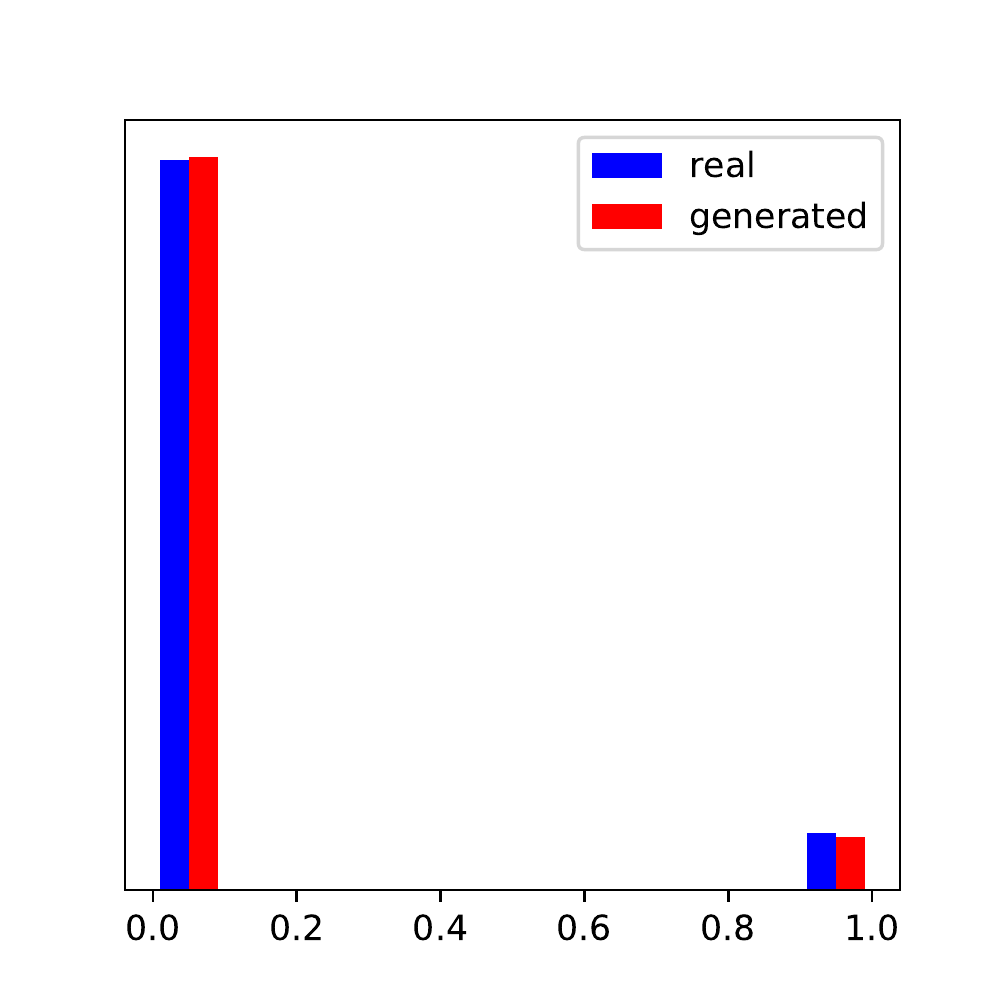}
\caption{Hispanic}
\end{subfigure} 
\begin{subfigure} [] {0.29\textwidth} 
\includegraphics[width=\textwidth]{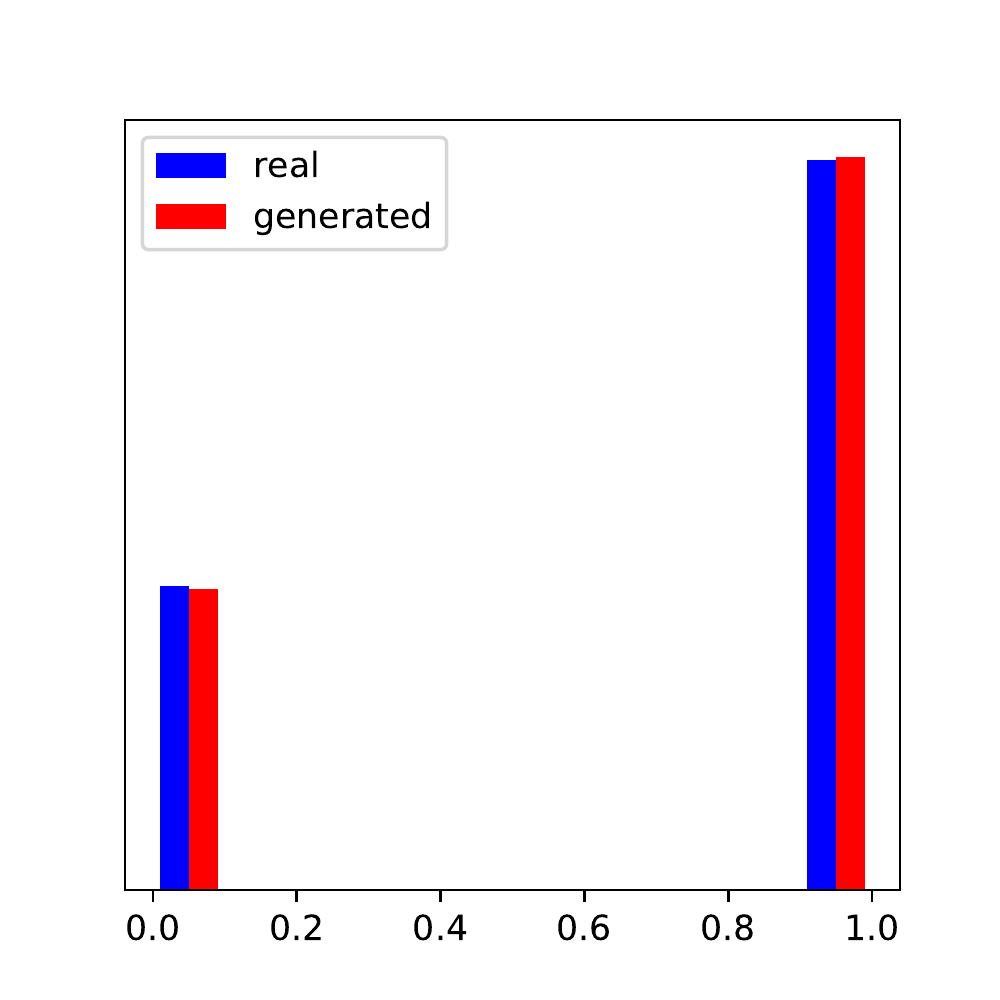}
\caption{Married}
\end{subfigure} 
\begin{subfigure} [] {0.29\textwidth} 
\includegraphics[width=\textwidth]{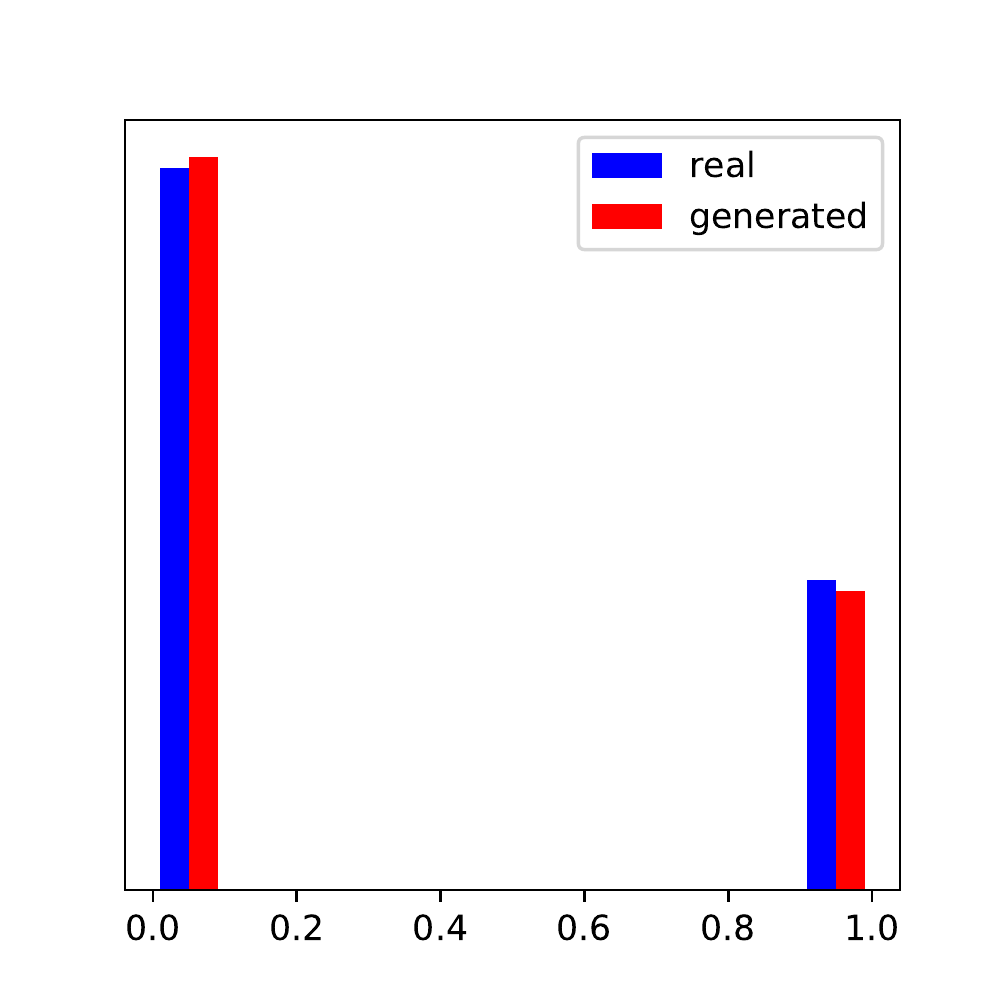}
\caption{No Degree}
\end{subfigure} 
\begin{subfigure} [] {0.29\textwidth} 
\includegraphics[width=\textwidth]{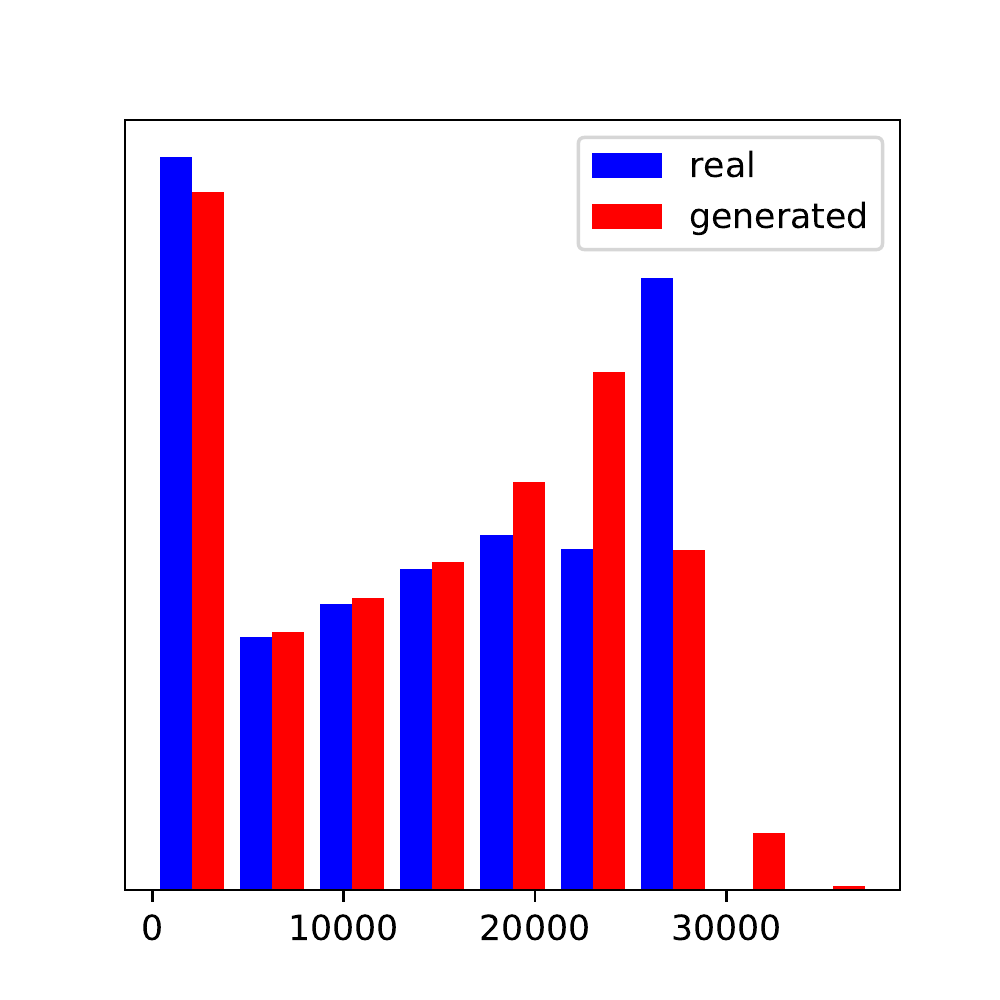}
\caption{Earnings 1974}
\end{subfigure} 
\begin{subfigure} [] {0.29\textwidth} 
\includegraphics[width=\textwidth]{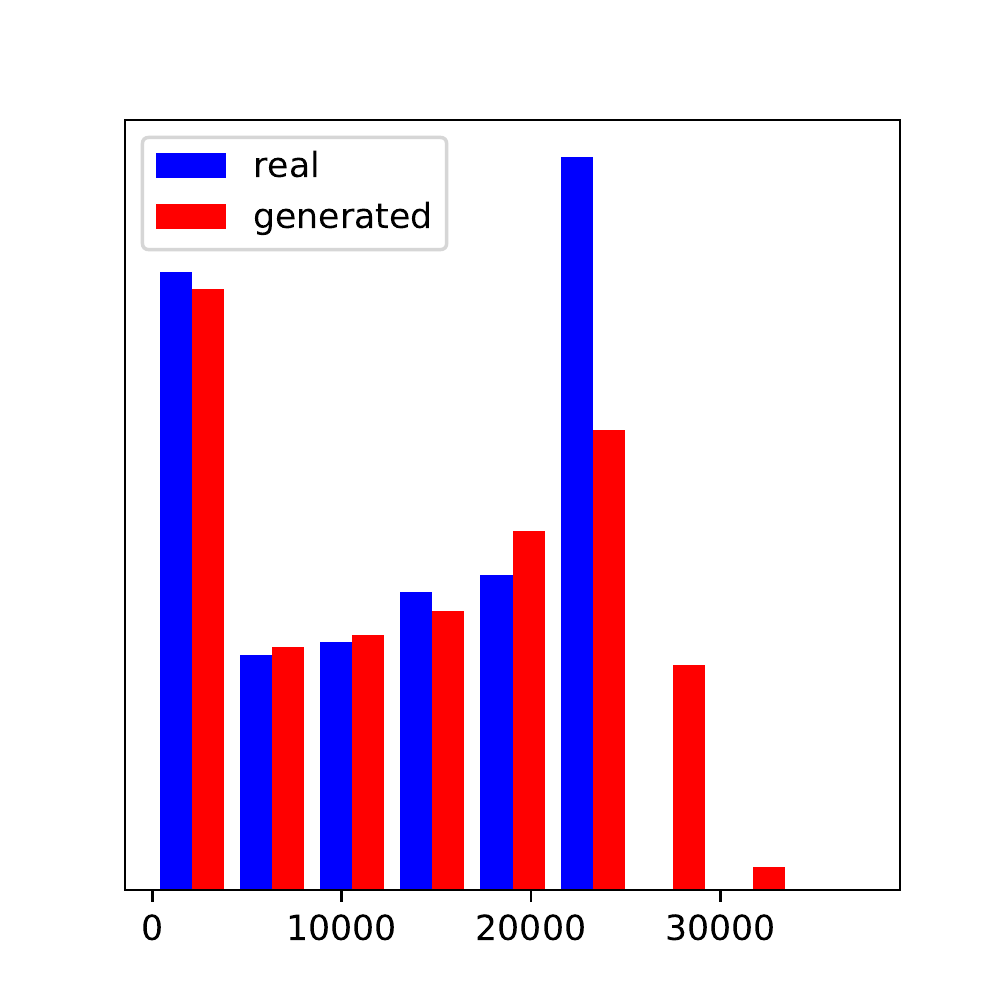}
\caption{Earnings 1975}
\end{subfigure} 
\begin{subfigure} [] {0.29\textwidth} 
\includegraphics[width=\textwidth]{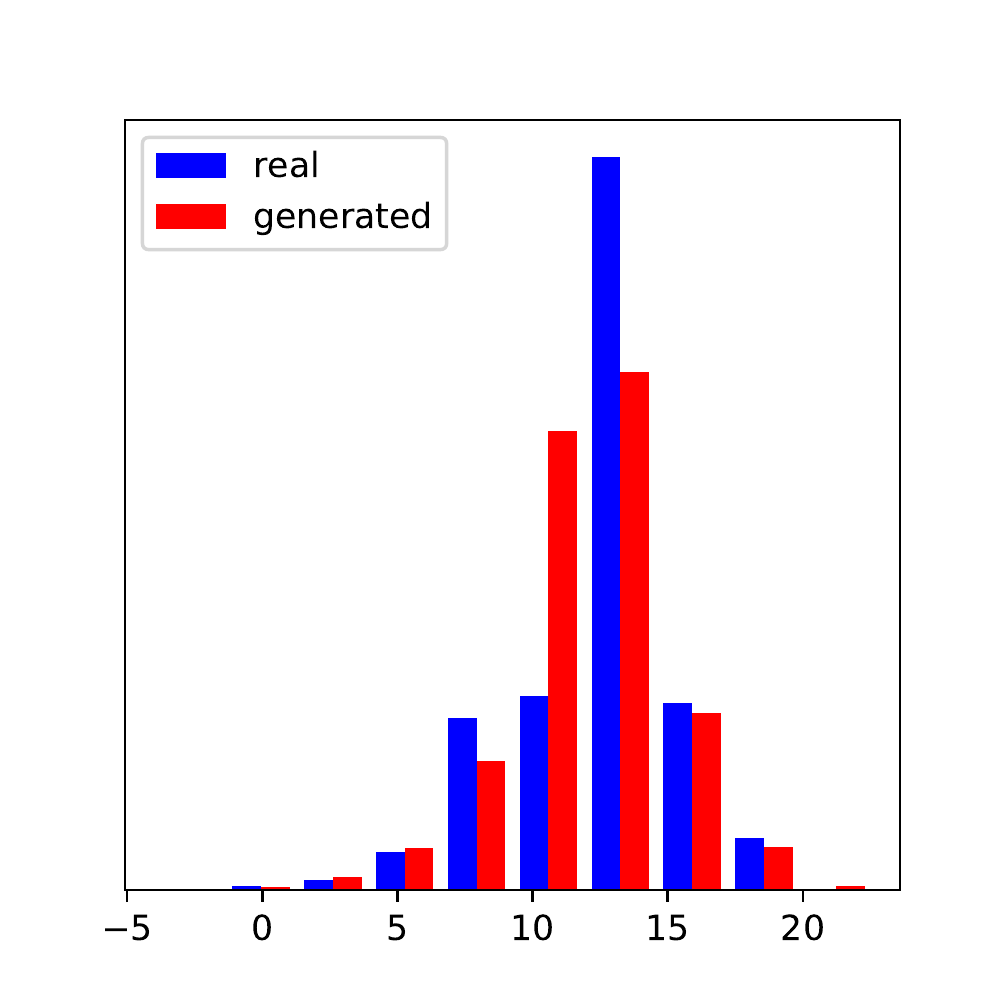}
\caption{Education}
\end{subfigure} 
\begin{subfigure} [] {0.29\textwidth} 
\includegraphics[width=\textwidth]{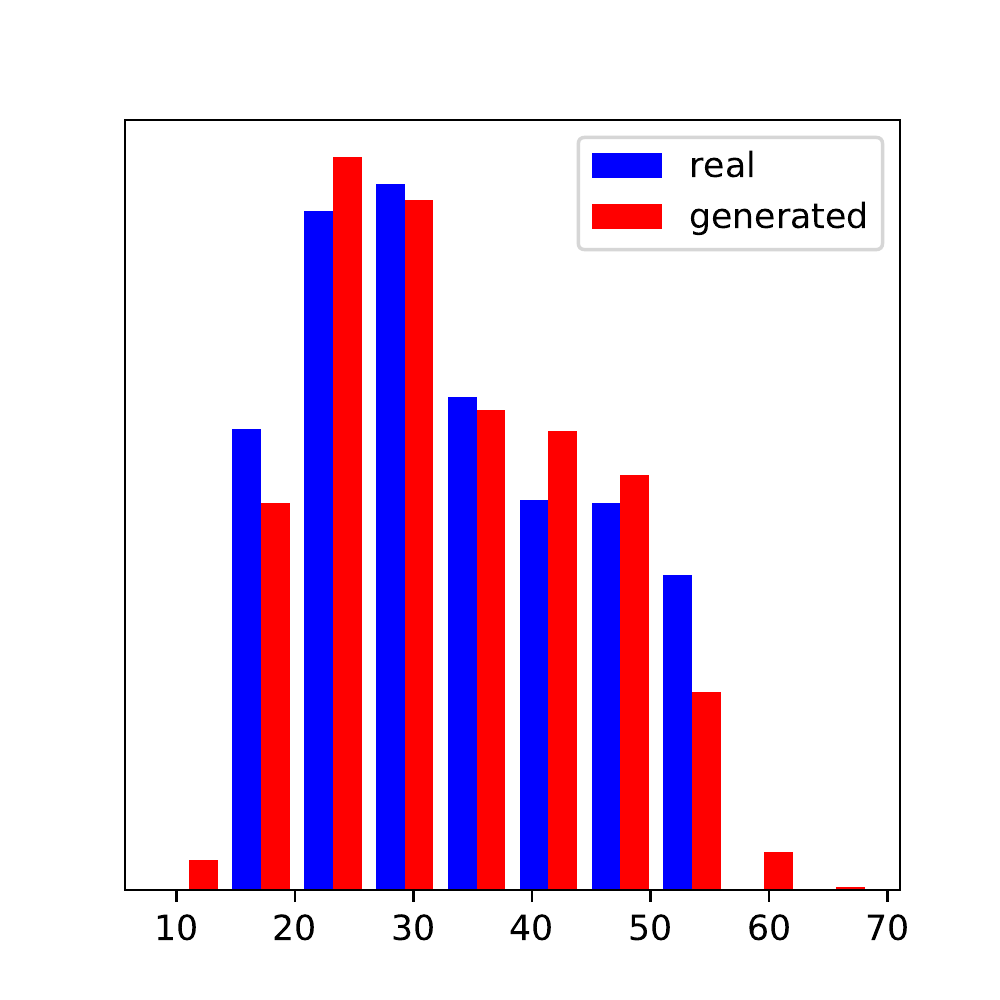}
\caption{Age}
\end{subfigure} 
\end{figure}

\begin{figure}[!ht]
\centering
\caption{\sc Correlations for CPS Data}
\includegraphics[width=6.5in]{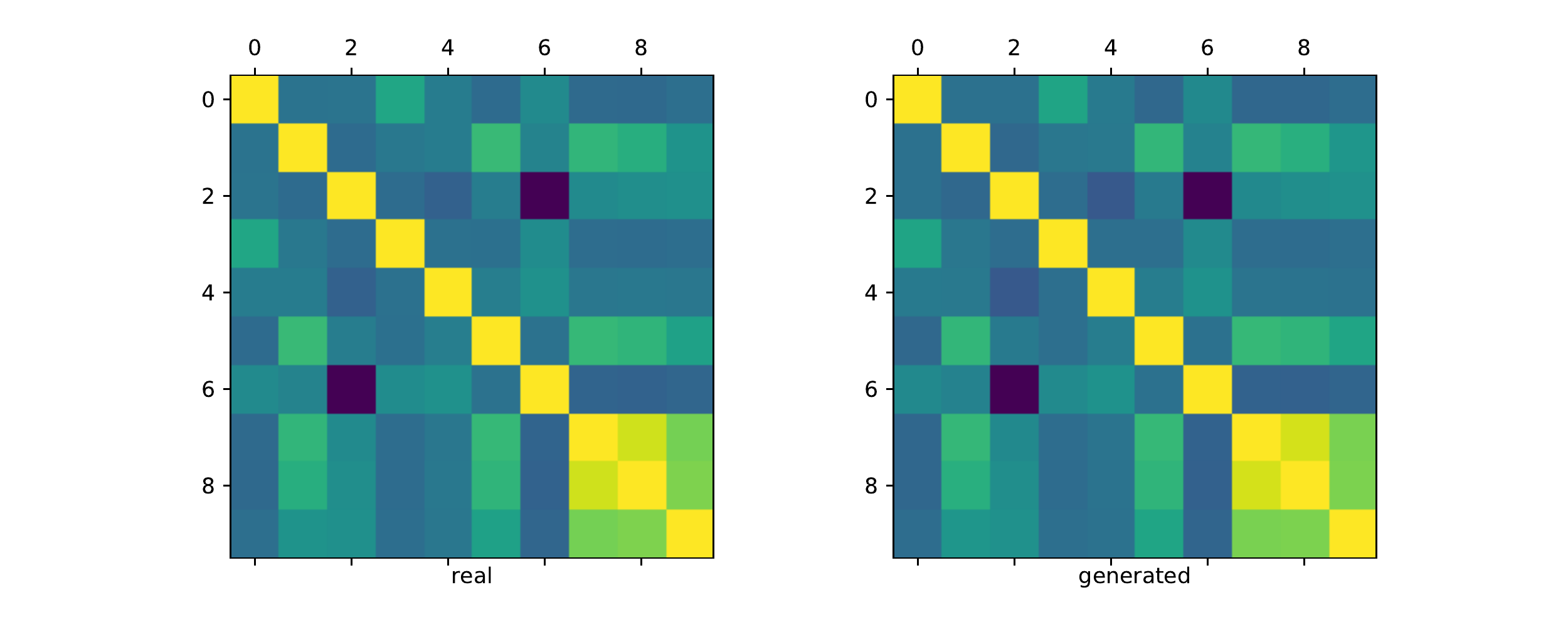}
\end{figure}

\begin{figure}[!ht]
\centering
\caption{\sc Conditional Histograms for CPS Data}
\begin{subfigure}[] {0.49\textwidth} 
\includegraphics[width=\textwidth]{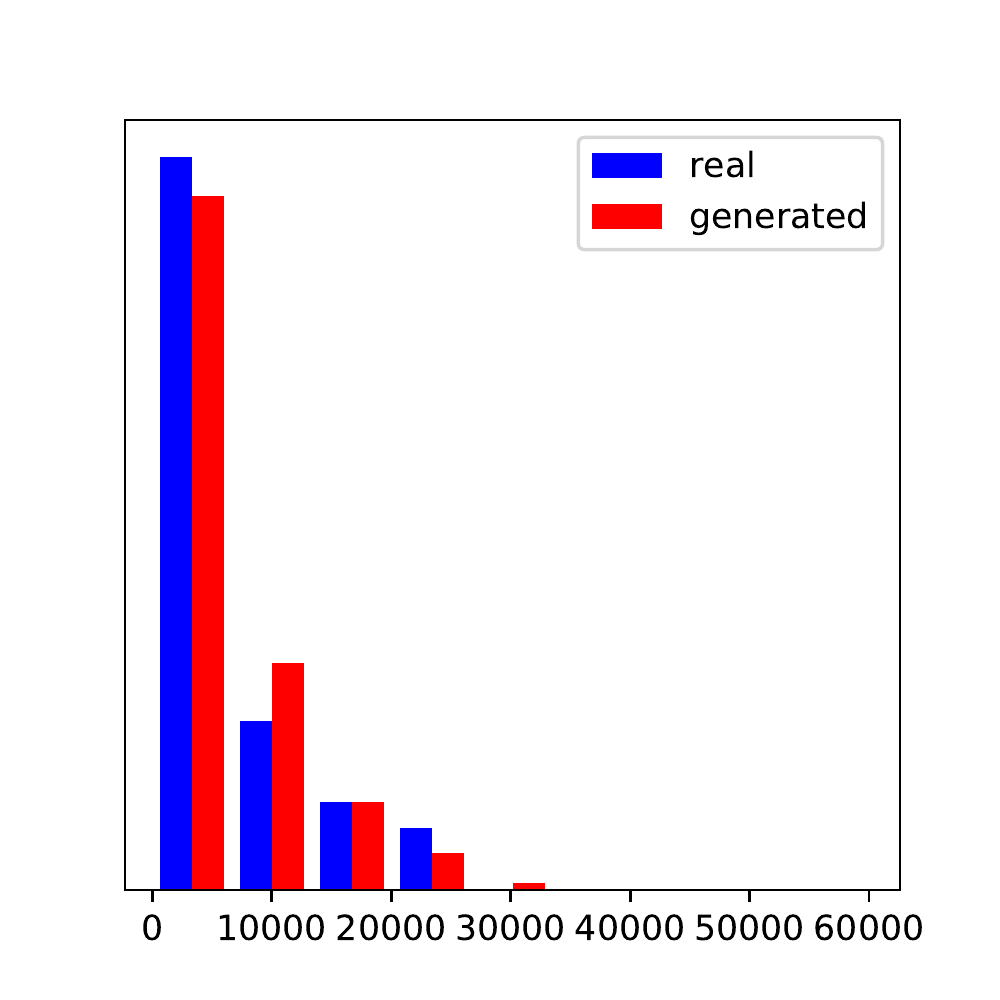}
\caption{Earnings 1978 $|$ Earnings 1974$ =0$} 
\end{subfigure} 
\begin{subfigure}[] {0.49\textwidth} 
\includegraphics[width=\textwidth]{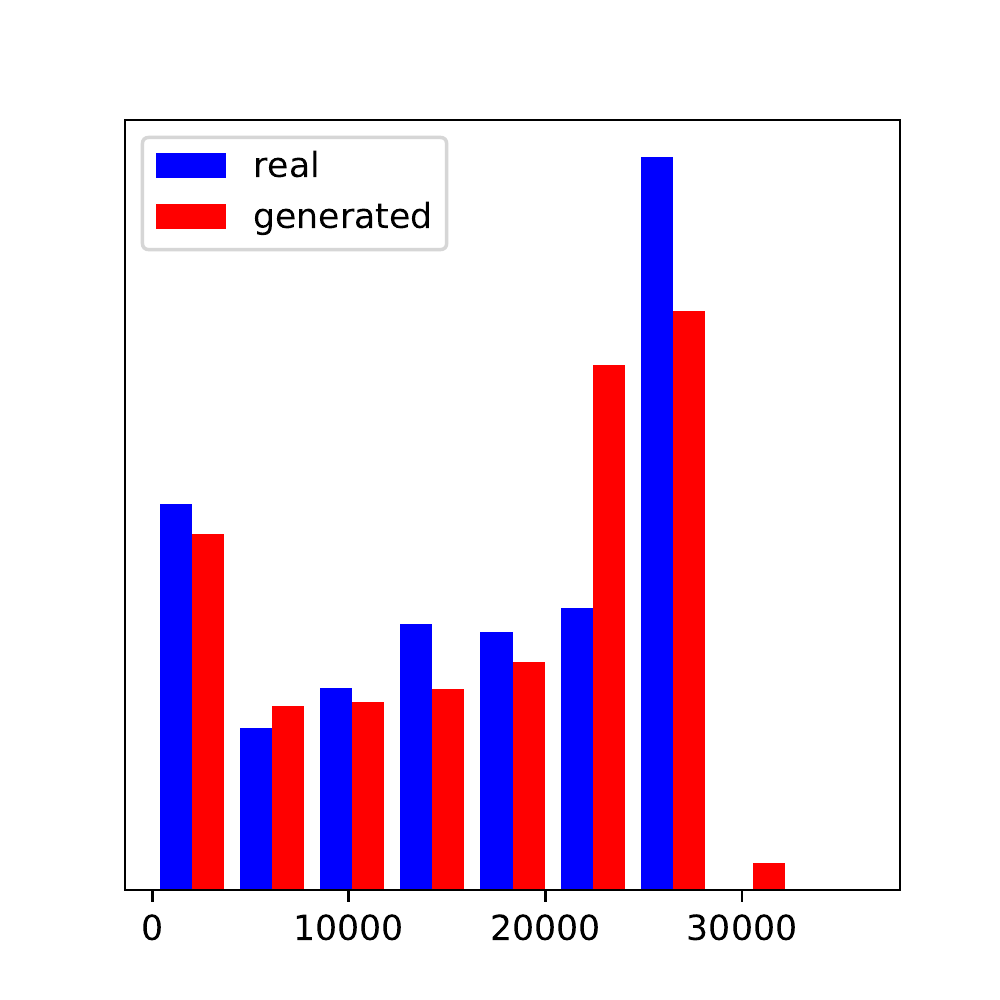}
\caption{Earnings 1978 $|$ Earnings 1974 $>$ 0} 
\end{subfigure} 
\end{figure}

In Table \ref{table::wd} we report the exact Wasserstein distances, calculated via linear programming, for both the GAN and a multivariate normal distribution fitted to the respective data sets below as a comprehensive measure of fit. The experimental and PSID datasets are averaged over 10 samples from the generator of equal size to the original dataset, and for the larger CPS data the reported distance is an average over 3 samples of equal size to the original dataset.  We find that the Wasserstein distance is considerably smaller for the WGAN simulations than for the multi-variate normal simulations, which just match the mean and variance of the real data. For larger datasets, calculating the exact Wasserstein distance via linear programming is infeasible, but instead an approximate version can be calculated by adding an entropic regularization term that smoothes the optimal transport problem necessary to calculate the Wasserstein distance, see \citet{cuturi2013sinkhorn}.  \\

\begin{table}[ht]
 \caption{\sc Wasserstein Distance Between Generated Data and Empirical Distribution}
 \label{table::wd} 
 \vskip1cm
\centering 
\begin{tabular}{lccc}
\toprule
Dataset &   GAN &  Multivariate Normal  &  Ratio \\
\midrule
    Experimental controls &             1606  &             4972  &   0.32 \\
CPS controls &            1629  &             5005 &   0.33 \\
   PSID controls &             3152 &             5586 &   0.56  \\
\bottomrule\\
\end{tabular}
\end{table}

Finally we assess how well the conditional expectation implied by the generator is approximated by a linear function. To do so we fit a linear model, a random forest and a neural net to this regression, and compare the goodness of fit out of sample in Table \ref{table2a}. We do so both with the actual data and the generated data. The out of sample $R^2$ is estimated using five-fold cross-validation. For the experimental generated data, which has a small sample size, the average cross-validated $R^2$ over 50 different samples from the generator is reported. We find that for the experimental, PSID and CPS samples the model fit is similar between the real and generated data, for all three models. Furthermore, the ranking of the three predictive models in terms of out of sample $R^2$ is similar between real and generated data. This suggests the generated data captures any non-linearity of the conditional expectation in the real data well. 

\begin{table}[t]
\caption{\sc Out-of-Sample Goodness of Fit  ($R^2$) on Real and Generated Data}
\centering
\vskip1cm 
\begin{tabular}{llccc}
\hline \hline \\ 
&& Experimental Controls & CPS Controls & PSID Controls \\ \hline  \\
Real Data& Linear Model & -0.04 & 0.47 & 0.56  \\
Real Data& Random Forest & -0.06 & 0.48  & 0.58  \\
Real Data& Neural Net & -0.10 & 0.48 &   0.55  \\
\\
Generated Data& Linear Model & 0.10 & 0.47 & 0.60  \\
Generated Data& Random Forest & 0.09 & 0.49 &  0.66 \\
Generated Data& Neural Net &   -0.07 &   0.50 &  0.60   \\
\\ \hline
\end{tabular}
\label{table2a}
\end{table}



\section{Comparing Estimators for Average Treatment Effects}
\label{section:comparing_estimators}

In this section we implement the WGANs to generate data sets to compare different estimators for the average treatment effects. We do this in three settings, first with the experimental LDW-E data, second with the LDW-CPS comparison group, and third with the LDW-PSID comparison group.

\subsection{Estimators}
\label{subsection:definitions}
We compare thirteen estimators for the average effect for the treated. Nine of them fit into a set where we compare three methods for estimating the two nuisance functions, the propensity score  $e(x)$ and the conditional outcome mean $\mu(0,x)$ (linear and logit models, random forests, and neural nets), with three ways of combining these estimates of the nuisance functions (difference in estimated conditional means, Horvitz-Thompson type inverse propensity score weighting, and double robust methods), and four are stand-alone estimators. All estimators that involve estimating the propensity score use trimming on the estimated propensity score,  dropping all observations with an estimated propensity score larger than 0.95. See
  \citet{crump2009dealing} for discussions of the importance of trimming in general.

  For estimating the two nuisance functions $e(x)$ and $\mu(0,x)$  we consider three methods:
\begin{enumerate}
\item A logit model for the  propensity score and a linear model for the conditional outcome mean, given the set of eight pre-treatment variables. Denote the estimator for the conditional mean by $\hat \mu^\lm(0,x)$, and the estimator for the propensity score by $\hat e^\lm(x)$. In settings with high dimensional covariates one might modify these estimators using regularization as in, for example, \citet{farrell2015robust}.
\item Random Forests for the propensity score and the conditional outcome mean. Denote the estimator for the conditional mean by $\hat \mu^\rf(0,x)$, and the estimator for the propensity score by $\hat e^\rf(x)$, {\it e.g.,} \citet{wager2018estimation}.  
\item Neural Nets  for the propensity score and the conditional outcome mean. Denote the estimator for the conditional mean by $\hat \mu^\nnn(0,x)$, and the estimator for the propensity score by $\hat e^\nnn(x)$. \citet{farrell2018deep} establish the convergence rates for deep neural networks necessary for semiparametric inference.
\end{enumerate}

We also consider three methods for incorporating the estimated nuisance functions into an estimator for the ATT:
\begin{enumerate}
\item Use the estimated conditional outcome mean by averaging the difference between the realized outcome and the esitmated control outcome, averaging this over the treated observations ({\it e.g.,} \citet{hahn1998role}:
\[ \hat\tau^\cm=\frac{1}{N_1}\sum_{i:W_i=1} \Bigl( Y_i-\hat\mu(0,X_i)\Bigr).\]
\item Use the estimated propensity score to weight the control observations, {\it e.g.,} \citet{hirano2003efficient}:
\[  \hat\tau^{\hoth}=\sum_{i} \left( \frac{W_i}{N_1} Y_i - (1-W_i) Y_i \frac{\hat e(X_i)}{1-\hat e(X_i)}\biggl/
\sum_{j=1}^N (1-W_j)\frac{\hat e(X_j)}{1-\hat e(X_j)}\right).\]
\item Use both the estimated conditional mean and the estimated propensity score in a double robust approach, {\it e.g.,} \citet{scharfstein1999comments, chernozhukov2017double}:
\[  \hat\tau^{\dr}=\sum_{i} \left( \frac{W_i}{N_1} \Bigl( Y_i-\hat\mu(0,X_i)\Bigr) - (1-W_i) \Bigl( Y_i-\hat\mu(0,X_i)\Bigr) \frac{\hat e(X_i)}{1-\hat e(X_i)}\biggl/
\sum_{j=1}^N (1-W_j)\frac{\hat e(X_j)}{1-\hat e(X_j)}\right).\]

\end{enumerate}
Note that for the neural net and the random forest implementation, we use sample splitting as in \cite{chernozhukov2017double}. We indicate the respective combination of nuisance function estimators and ATT formulas by combining the superscripts, yielding the nine estimators $\hat\tau^{\cm,\lm}$, $\hat\tau^{\cm,\rf}$, $\hat\tau^{\cm,\nnn}$, $\hat\tau^{\hoth,\lm}$ , $\hat\tau^{\hoth,\rf}$, $\hat\tau^{\hoth,\nnn}$, $\hat\tau^{\dr,\lm}$, $\hat\tau^{\dr,\rf}$ and $\hat\tau^{\dr,\nnn}$.

\subsection{Estimates for LDW Data}

First we compute all thirteen estimators on the three samples for the original LDW data. The results are reported in Table \ref{tabel3}.

\begin{table}[ht]
 \caption{\sc Estimates Based on LDW Data}
 \vskip1cm
 \begin{center}
\begin{tabular}{lrrrrrr}
\toprule
\multicolumn{1}{c}{ } & \multicolumn{2}{c}{Experimental} & \multicolumn{2}{c}{CPS} & \multicolumn{2}{c}{PSID} \\
\cmidrule(l{3pt}r{3pt}){2-3} \cmidrule(l{3pt}r{3pt}){4-5} \cmidrule(l{3pt}r{3pt}){6-7}
  & estimate & s.e. & estimate & s.e. & estimate & s.e.\\
\midrule
\addlinespace[0.3em]
\multicolumn{7}{l}{\textbf{Baselines}}\\
\hspace{1em}DIFF & 1.79 & 0.63 & -8.50 & 0.71 & -15.20 & 1.15\\
\hspace{1em}BCM & 2.12 & 0.88 & 2.15 & 0.87 & 0.57 & 1.47\\
\addlinespace[0.3em]
\multicolumn{7}{l}{\textbf{Outcome Models}}\\
\hspace{1em}L & 1.79 & 0.57 & 0.69 & 0.60 & 0.79 & 0.60\\
\hspace{1em}RF & 1.69 & 0.58 & 0.85 & 0.60 & -0.20 & 0.56\\
\hspace{1em}NN & 1.49 & 0.59 & 1.70 & 0.60 & 1.47 & 0.60\\
\addlinespace[0.3em]
\multicolumn{7}{l}{\textbf{Propensity Score Models}}\\
\hspace{1em}L & 1.81 & 0.83 & 1.18 & 0.77 & 1.26 & 1.13\\
\hspace{1em}RF & 1.90 & 0.86 & 0.73 & 0.82 & 0.24 & 1.00\\
\hspace{1em}NN & 1.69 & 0.86 & 1.38 & 0.77 & 0.42 & 1.45\\
\addlinespace[0.3em]
\multicolumn{7}{l}{\textbf{Doubly Robust Methods}}\\
\hspace{1em}L & 1.80 & 0.67 & 1.27 & 0.65 & 1.50 & 0.97\\
\hspace{1em}RF & 1.93 & 0.70 & 1.63 & 0.76 & 0.98 & 0.83\\
\hspace{1em}NN & 1.90 & 0.75 & 1.63 & 0.72 & 1.56 & 0.76\\
\hspace{1em}CF & 1.72 & 0.68 & 1.58 & 0.67 & 0.59 & 0.78\\
\hspace{1em}RB & 1.73 & 0.70 & 0.93 & 0.62 & 0.72 & 0.79\\
\bottomrule
\end{tabular}
 \end{center}
 \label{tabel3}
\end{table}

\subsection{Simulation Results for the Experimental Control Sample}

Next, we report results for the comparison of all the estimators for the experimental sample in Table \ref{tabelexp}. 
We draw 2,000 samples from the population (the sample of size 1,000,000) and calculate the estimated treatment effect for each sample and each of the thirteen estimators. The population value of the treatment effect is the average treatment effect for treated in the generated population of 1 million individuals. 
We report the average bias of each estimator across the 2,000 samples, the standard deviation for each estimator across the 2,000 samples, the root-mean-squared error (RMSE) and the coverage rates over the 2,000 replications. 

For the experimental sample, the RMSEs for the different estimators are fairly similar, ranging from 0.49 (for the residual balancing estimator) to 1.32 for the outcome model based on neural nets.  Because there is balance between the treatment and control group due to random assignment of the treatment, it is not surprising that all methods perform fairly well.  The double robust methods do particularly well in terms of coverage rates for the 95\% confidence intervals.

\begin{table}[H]
\caption{\sc Estimates Based on LDW Experimental Data (2,000 Replications)}
\label{tabelexp} 
\centering\begin{table}[H]
\centering
\begin{tabular}{lrrrr}
\toprule
method & rmse & bias & sdev & coverage\\
\midrule
\addlinespace[0.3em]
\multicolumn{5}{l}{\textbf{Baselines}}\\
\hspace{1em}DIFF & 0.49 & 0.06 & 0.48 & 0.94\\
\hspace{1em}BCM & 0.58 & 0.00 & 0.58 & 0.96\\
\addlinespace[0.3em]
\multicolumn{5}{l}{\textbf{Outcome Models}}\\
\hspace{1em}L & 0.52 & -0.06 & 0.51 & 0.88\\
\hspace{1em}RF & 0.51 & -0.07 & 0.50 & 0.88\\
\hspace{1em}NN & 1.32 & 0.04 & 1.32 & 0.75\\
\addlinespace[0.3em]
\multicolumn{5}{l}{\textbf{Propensity Score Models}}\\
\hspace{1em}L & 0.52 & -0.08 & 0.52 & 0.99\\
\hspace{1em}RF & 0.52 & -0.06 & 0.51 & 0.99\\
\hspace{1em}NN & 0.52 & 0.01 & 0.52 & 0.99\\
\addlinespace[0.3em]
\multicolumn{5}{l}{\textbf{Doubly Robust Methods}}\\
\hspace{1em}L & 0.51 & -0.08 & 0.51 & 0.95\\
\hspace{1em}RF & 0.52 & -0.04 & 0.52 & 0.95\\
\hspace{1em}NN & 0.79 & -0.05 & 0.79 & 0.95\\
\hspace{1em}CF & 0.50 & -0.09 & 0.49 & 0.94\\
\hspace{1em}RB & 0.52 & -0.09 & 0.51 & 0.95\\
\bottomrule
\end{tabular}
\end{table}
\end{table}

\subsection{Simulation Results for the CPS Control Sample}

Next, we report results for the comparison of the twelve estimators for the CPS comparison sample in Table \ref{tabelcps}.
As expected, given the substantial differences in characteristics between the treatment group and the control group, in this exercise we find considerably bigger differences in the performances of the different estimators.  The  double robust methods generally do well here. The biases for some of the estimators that are not doubly robust  are substantial, contributing to their confidence intervals having poor coverage rates.

\begin{table}[H]
\caption{\sc Estimates Based on LDW-CPS Data  (2,000 Replications)}
\label{tabelcps} 
\centering\begin{table}[H]
\centering
\begin{tabular}{lrrrr}
\toprule
method & rmse & bias & sdev & coverage\\
\midrule
\addlinespace[0.3em]
\multicolumn{5}{l}{\textbf{Baselines}}\\
\hspace{1em}DIFF & 11.12 & -11.11 & 0.45 & 0.00\\
\hspace{1em}BCM & 0.73 & 0.07 & 0.73 & 0.96\\
\addlinespace[0.3em]
\multicolumn{5}{l}{\textbf{Outcome Models}}\\
\hspace{1em}L & 2.14 & -2.08 & 0.51 & 0.02\\
\hspace{1em}RF & 1.00 & -0.87 & 0.51 & 0.54\\
\hspace{1em}NN & 0.63 & 0.14 & 0.61 & 0.88\\
\addlinespace[0.3em]
\multicolumn{5}{l}{\textbf{Propensity Score Models}}\\
\hspace{1em}L & 0.51 & 0.00 & 0.51 & 0.98\\
\hspace{1em}RF & 1.00 & -0.87 & 0.50 & 0.73\\
\hspace{1em}NN & 0.65 & 0.23 & 0.61 & 0.94\\
\addlinespace[0.3em]
\multicolumn{5}{l}{\textbf{Doubly Robust Methods}}\\
\hspace{1em}L & 0.53 & 0.03 & 0.53 & 0.96\\
\hspace{1em}RF & 0.54 & -0.05 & 0.54 & 0.93\\
\hspace{1em}NN & 0.62 & 0.20 & 0.58 & 0.94\\
\hspace{1em}CF & 0.55 & 0.11 & 0.53 & 0.91\\
\hspace{1em}RB & 0.57 & -0.22 & 0.52 & 0.89\\
\bottomrule
\end{tabular}
\end{table}
\end{table}

\subsection{Simulation Results for the PSID Control Sample}

Third, we report results for the comparison of the twelve estimators for the psid comparison sample in Table \ref{tabelpsid}. Again the double robust methods do well overall. Note that the linear methods do particularly well in terms of bias. 

\begin{table}[H]
\caption{\sc Estimates Based on LDW-PSID Data (2,000 Replications)}
\label{tabelpsid} 
\centering\begin{table}[H]
\centering
\begin{tabular}{lrrrr}
\toprule
method & rmse & bias & sdev & coverage\\
\midrule
\addlinespace[0.3em]
\multicolumn{5}{l}{\textbf{Baselines}}\\
\hspace{1em}DIFF & 18.81 & -18.81 & 0.53 & 0.00\\
\hspace{1em}BCM & 0.98 & -0.02 & 0.98 & 0.98\\
\addlinespace[0.3em]
\multicolumn{5}{l}{\textbf{Outcome Models}}\\
\hspace{1em}L & 1.95 & -1.82 & 0.72 & 0.12\\
\hspace{1em}RF & 2.30 & -2.22 & 0.62 & 0.02\\
\hspace{1em}NN & 2.97 & -0.93 & 2.82 & 0.59\\
\addlinespace[0.3em]
\multicolumn{5}{l}{\textbf{Propensity Score Models}}\\
\hspace{1em}L & 1.11 & -0.64 & 0.91 & 0.96\\
\hspace{1em}RF & 2.21 & -2.05 & 0.82 & 0.32\\
\hspace{1em}NN & 1.82 & -1.43 & 1.11 & 0.69\\
\addlinespace[0.3em]
\multicolumn{5}{l}{\textbf{Doubly Robust Methods}}\\
\hspace{1em}L & 0.98 & -0.35 & 0.92 & 0.94\\
\hspace{1em}RF & 0.98 & -0.57 & 0.80 & 0.84\\
\hspace{1em}NN & 0.98 & -0.38 & 0.90 & 0.92\\
\hspace{1em}CF & 1.13 & -0.89 & 0.69 & 0.73\\
\hspace{1em}RB & 1.06 & 0.33 & 1.01 & 0.75\\
\bottomrule
\end{tabular}
\end{table}
\end{table}

\section{Robustness of the Simulations}

The algorithm developed in this paper leads, for a given data set, to a RMSE for each estimator, and, based on that, a unique ranking of a set of estimators. However, it does not come with a measure of robustness of that ranking or the RMSE it is based on. The estimated RMSEs and the implied ranking of the estimators could be different if we change the set up. In particular we may be concerned with the robustness of the bias component of the RMSE. 
In this section we discuss a number approaches to assessing how robust the rankings are.

\subsection{Robustness to Sample}
 
We apply the WGANs to $M=10$ samples drawn without replacement  from the original sample. Each sample is 80\% of the size of the original sample. We use these subsamples to train a WGAN and for each WGAN, draw 10,000 samples from the population distribution and calculate RMSE, bias, standard deviation, coverage, and power. The main question of interest is by how much the results vary across the different subsamples of the data. The table gives the average of each metric of interest, calculated across the 10 different synthetic populations trained from  the 10 different subsamples of original data. The averages are close to the point estimates of the metrics from the full sample. We also show the standard deviation; although there is substantial variation in the estimates over the synthetic populations trained on different 80\% subsamples of the dataset, the conclusion that the doubly-robust methods generally outperform the other methods still holds.
 
\begin{table}[H]
\caption{\sc Robustness of Ranking for LDW-CPS, Average and Standard Deviations of Metrics over $M=10$ Samples Drawn from Original Sample}
\label{tablecv} 
\centering
\vskip1cm
\begin{tabular}{lrrrr}
\toprule
method & rmse & bias & sdev & coverage\\
\midrule
\addlinespace[0.3em]
\multicolumn{5}{l}{\textbf{Baselines}}\\
\hspace{1em}DIFF & 10.12 (1.29) & -10.11 (1.29) & 0.45 (0.04) & 0.00 (0.00)\\
\hspace{1em}BCM & 0.78 (0.13) & -0.04 (0.13) & 0.77 (0.12) & 0.96 (0.02)\\
\addlinespace[0.3em]
\multicolumn{5}{l}{\textbf{Outcome Models}}\\
\hspace{1em}L & 1.09 (0.48) & -0.69 (0.88) & 0.49 (0.04) & 0.50 (0.35)\\
\hspace{1em}RF & 0.87 (0.32) & -0.59 (0.52) & 0.51 (0.04) & 0.64 (0.29)\\
\hspace{1em}NN & 0.70 (0.16) & 0.14 (0.39) & 0.60 (0.04) & 0.82 (0.09)\\
\addlinespace[0.3em]
\multicolumn{5}{l}{\textbf{Propensity Score Models}}\\
\hspace{1em}L & 0.75 (0.30) & 0.07 (0.63) & 0.52 (0.04) & 0.89 (0.16)\\
\hspace{1em}RF & 0.98 (0.36) & -0.76 (0.52) & 0.50 (0.04) & 0.73 (0.27)\\
\hspace{1em}NN & 0.68 (0.12) & 0.04 (0.28) & 0.63 (0.05) & 0.96 (0.03)\\
\addlinespace[0.3em]
\multicolumn{5}{l}{\textbf{Doubly Robust Methods}}\\
\hspace{1em}L & 0.75 (0.32) & 0.14 (0.63) & 0.53 (0.05) & 0.82 (0.19)\\
\hspace{1em}RF & 0.62 (0.11) & 0.05 (0.33) & 0.54 (0.05) & 0.89 (0.05)\\
\hspace{1em}NN & 0.65 (0.12) & 0.07 (0.23) & 0.62 (0.07) & 0.94 (0.03)\\
\hspace{1em}CF & 0.67 (0.16) & 0.14 (0.39) & 0.56 (0.06) & 0.83 (0.08)\\
\hspace{1em}RB & 0.82 (0.24) & 0.00 (0.70) & 0.53 (0.03) & 0.69 (0.19)\\
\bottomrule
\end{tabular}
\end{table}

\subsection{Robustness to Model Architecture}

 We also investigate the robustness to the architecture of the critic and generator, within a similar complexity class of neural networks. Recall that the 
 architecture of the generator and critic both have three hidden layers, with dimensions
 $(d_X+M,128), $ $(128,128)$ and $(128,128)$. 
 The first alternative architecture (Alt1) considered has a generator hidden layer with dimensions $[64,128,256]$ and a critic hidden layer with dimensions $[256,128,64]$. The second alternative architecture (Alt2)  considered has a generator hidden layer with dimensions $[128,256,64]$ and a critic hidden layer with dimensions $[64,256,128]$. We do not find that our results are overly sensitive to a certain WGAN architecture. We find that the RMSE, bias, and standard deviation estimates or each treatment effect estimator are similar for the main and two alternative specifications.

\begin{table}[H]
\caption{\sc Robustness  to Model Architecture for LDW-CPS }
\label{tablearch} 
\centering\begin{table}[H]
\centering
\begin{tabular}{lrrrrrrrrr}
\toprule
\multicolumn{1}{c}{ } & \multicolumn{3}{c}{rmse} & \multicolumn{3}{c}{bias} & \multicolumn{3}{c}{sdev} \\
\cmidrule(l{3pt}r{3pt}){2-4} \cmidrule(l{3pt}r{3pt}){5-7} \cmidrule(l{3pt}r{3pt}){8-10}
method & Main & Alt1 & Alt2 & Main & Alt1 & Alt2 & Main & Alt1 & Alt2\\
\midrule
\addlinespace[0.3em]
\multicolumn{10}{l}{\textbf{Baselines}}\\
\hspace{1em}DIFF & 11.12 & 9.80 & 11.50 & -11.11 & -9.79 & -11.49 & 0.45 & 0.43 & 0.45\\
\hspace{1em}BCM & 0.73 & 0.66 & 0.57 & 0.07 & 0.07 & 0.03 & 0.73 & 0.65 & 0.57\\
\addlinespace[0.3em]
\multicolumn{10}{l}{\textbf{Outcome Models}}\\
\hspace{1em}L & 2.14 & 0.70 & 2.14 & -2.08 & -0.52 & -2.08 & 0.51 & 0.46 & 0.48\\
\hspace{1em}RF & 1.00 & 0.72 & 1.38 & -0.87 & -0.55 & -1.30 & 0.51 & 0.46 & 0.47\\
\hspace{1em}NN & 0.63 & 0.54 & 0.73 & 0.14 & -0.03 & -0.45 & 0.61 & 0.54 & 0.57\\
\addlinespace[0.3em]
\multicolumn{10}{l}{\textbf{Propensity Score Models}}\\
\hspace{1em}L & 0.51 & 0.51 & 1.15 & 0.00 & -0.19 & -1.04 & 0.51 & 0.48 & 0.48\\
\hspace{1em}RF & 1.00 & 0.80 & 1.50 & -0.87 & -0.65 & -1.42 & 0.50 & 0.47 & 0.47\\
\hspace{1em}NN & 0.65 & 0.54 & 0.65 & 0.23 & -0.03 & -0.34 & 0.61 & 0.54 & 0.56\\
\addlinespace[0.3em]
\multicolumn{10}{l}{\textbf{Doubly Robust Methods}}\\
\hspace{1em}L & 0.53 & 0.50 & 1.11 & 0.03 & -0.15 & -0.98 & 0.53 & 0.48 & 0.50\\
\hspace{1em}RF & 0.54 & 0.51 & 0.64 & -0.05 & 0.08 & -0.40 & 0.54 & 0.51 & 0.50\\
\hspace{1em}NN & 0.62 & 0.54 & 0.54 & 0.20 & 0.06 & -0.16 & 0.58 & 0.54 & 0.52\\
\hspace{1em}CF & 0.55 & 0.53 & 0.52 & 0.11 & 0.20 & -0.16 & 0.53 & 0.49 & 0.49\\
\hspace{1em}RB & 0.57 & 0.50 & 0.97 & -0.22 & 0.09 & -0.82 & 0.52 & 0.49 & 0.52\\
\bottomrule
\end{tabular}
\end{table}
\end{table}

\subsection{Robustness to Size of Training Data} 

Next we change the size of the training sample to some fraction of the original sample. This is likely to make the generator more smooth because it has fewer data to be trained on. We still generate samples from the generator that are the same size as the original sample. The results are in Table  
\ref{tablesize}.

\begin{table}[H]
\caption {\sc RMSE for Estimators on LDW-CPS for Different Training Data Sizes}
\label{tablesize}
\centering\begin{table}[H]
\centering
\begin{tabular}{lrrrrrrrrr}
\toprule
Fraction of original sample & 0.2 & 0.3 & 0.4 & 0.5 & 0.6 & 0.7 & 0.8 & 0.9 & 1.0\\
\midrule
\addlinespace[0.3em]
\multicolumn{10}{l}{\textbf{Baselines}}\\
\hspace{1em}DIFF & 10.17 & 10.84 & 9.88 & 10.09 & 11.10 & 10.45 & 10.41 & 10.55 & 11.12\\
\hspace{1em}BCM & 0.60 & 0.73 & 0.64 & 0.72 & 0.78 & 0.71 & 0.72 & 0.65 & 0.73\\
\addlinespace[0.3em]
\multicolumn{10}{l}{\textbf{Outcome Models}}\\
\hspace{1em}L & 1.87 & 0.70 & 0.48 & 0.46 & 1.37 & 1.31 & 0.52 & 1.52 & 2.14\\
\hspace{1em}RF & 0.62 & 0.62 & 0.73 & 0.84 & 0.94 & 0.97 & 0.61 & 1.28 & 1.00\\
\hspace{1em}NN & 0.54 & 0.74 & 0.58 & 0.60 & 0.68 & 0.59 & 0.54 & 0.79 & 0.63\\
\addlinespace[0.3em]
\multicolumn{10}{l}{\textbf{Propensity Score Models}}\\
\hspace{1em}L & 0.74 & 0.66 & 0.49 & 0.76 & 0.60 & 0.53 & 0.49 & 1.37 & 0.51\\
\hspace{1em}RF & 0.94 & 0.71 & 0.69 & 1.06 & 1.10 & 1.08 & 0.89 & 1.49 & 1.00\\
\hspace{1em}NN & 0.54 & 0.71 & 0.57 & 0.59 & 0.68 & 0.56 & 0.61 & 0.59 & 0.65\\
\addlinespace[0.3em]
\multicolumn{10}{l}{\textbf{Doubly Robust Methods}}\\
\hspace{1em}L & 0.69 & 0.67 & 0.49 & 0.72 & 0.60 & 0.54 & 0.49 & 1.34 & 0.53\\
\hspace{1em}RF & 0.48 & 0.70 & 0.49 & 0.53 & 0.63 & 0.53 & 0.49 & 0.70 & 0.54\\
\hspace{1em}NN & 0.51 & 0.68 & 0.57 & 0.58 & 0.67 & 0.57 & 0.58 & 0.55 & 0.62\\
\hspace{1em}CF & 0.49 & 0.69 & 0.54 & 0.50 & 0.66 & 0.53 & 0.53 & 0.61 & 0.55\\
\hspace{1em}RB & 1.41 & 0.82 & 0.52 & 0.49 & 0.62 & 0.56 & 0.50 & 1.20 & 0.57\\
\bottomrule
\end{tabular}
\end{table}
\end{table}

\section{Imposing Restrictions}

So far the setting has been a just-identified one, so that the question is to generate data that mimic an actual data set without any restrictions. In many cases, however, we wish to generate data from a restricted set of distributions. Here we discuss a simple example to demonstrate how one can extend the ideas discussed so far to that case.

To make it specific, suppose we have two variables $(X_i,Y_i)$, and the model implies that $\mu(x)=\mathbb{E}[Y_i|X_i=x]$ is monotone (say, increasing) in $x$. This type of shape restriction is fairly common in structural models, for example, demand functions are typically monotone in prices. More subtle versions of that come up in auction models with the hazard rate of the bid distribution decreasing. So, the question is how to simulate data that look like the actual data, but imposing monotonicity of the conditional mean (or some other restriction). Here is one possible approach. Suppose we have a test statistic $T((X_1,Y_1),\ldots,(X_M,Y_M))$ that gives us a consistent test for the null hypothesis that the regression function is non-decreasing. For the generated noise $Z_1,\ldots,Z_M$, and the parameter of the generator $\theta_g$, we can write the statistic as a function of $\theta_g$ as $T(\theta_g)=T(g(Z_1,\theta_g),\ldots g(Z_M,\theta_g))$. Then, when we do gradient descent to find a new value for the generator parameter, we can add a penalty term $\lambda T(\theta_g)$ to the objective function so that we penalize changes of $\theta_g$ in the direction that increase the value of the test statistic. All this requires is a test statistic, and a value for the penalty term. 
To illustrate this we use a recent test statistic for monotonicity of the regression function proposed in \citet{chetverikov2019testing}, see also \citet{chetverikov2018econometrics}. We use the CPS subset of the LDW data and look at the case where $Y_i$ is earnings in 1978 and $X_i$ is the age. Here the regression function is far from montone in the actual sample, so we can see how imposing monotonicity changes the joint distribution.

To be more specific, our data for a particular batch are $(X_1,Y_1),\ldots,(X_M,Y_M)$, where $(X_i,Y_i)=g(Z_i,\theta_g)$. First we describe the calculation of the test statistic as a function 
\[ T=T((X_1,Y_1),\ldots,(X_M,Y_M)),\]
which implicitly defines it as a function of $\theta$ given the noise variables.
First, we need a variance estimator for $\mathbb{V}(Y_i|X_i=x)$, evaluated at the sample points $X_1,\ldots,X_M$. Suppose the observations are ordered, so that $X_i\leq X_{i+1}$ for all $i$.
Let \[ \hat\sigma_i^2=(Y_{i+1}-Y_i)^2/2.\]
(and $\hat\sigma_M^2=(Y_{M}-Y_{M-1})^2/2$.)
Define
\[ K(x)=0.75 (1-x^2),\]
\[ Q(x_1,x_2,x,h)=K((x_1-x)/h)K((x_2-x)/h),\]
\[ h_{\max}=\max_{1\leq i,j\leq M} |X_i-X_j|/2,\]
\[h_{\min}=h_{\max}(0.3/M^{0.95})^{1/3}.\]
\[ H_M=\{h_{\max},h_{\max}(0.5),h_{\max}(0.5)^2,\ldots,h_{\min}\}\]
Then
define
\[ b(x,h)=(1/2)\sum_{1\leq i,j\leq M} (Y_i-Y_j){\rm sign}(X_j-X_i)Q(X_i,X_j,x,h).\]
\[V(x,h)=\sum_{1\leq i\leq M}\hat\sigma^2_i \sum_{1\leq j\leq M} {\rm sign}(X_j-X_i) Q(X_i,X_j,x,h).\]
Then the test statistic is 
\[ T=T((X_1,Y_1),\ldots,(X_M,Y_M))=\max_{h\in H_M}\max_{1\leq i\leq M} \frac{b(X_i,h)}{V(X_i,h)}.\]

\begin{figure}[h!]
\includegraphics[trim=30 0 30 0,clip,width=\textwidth]{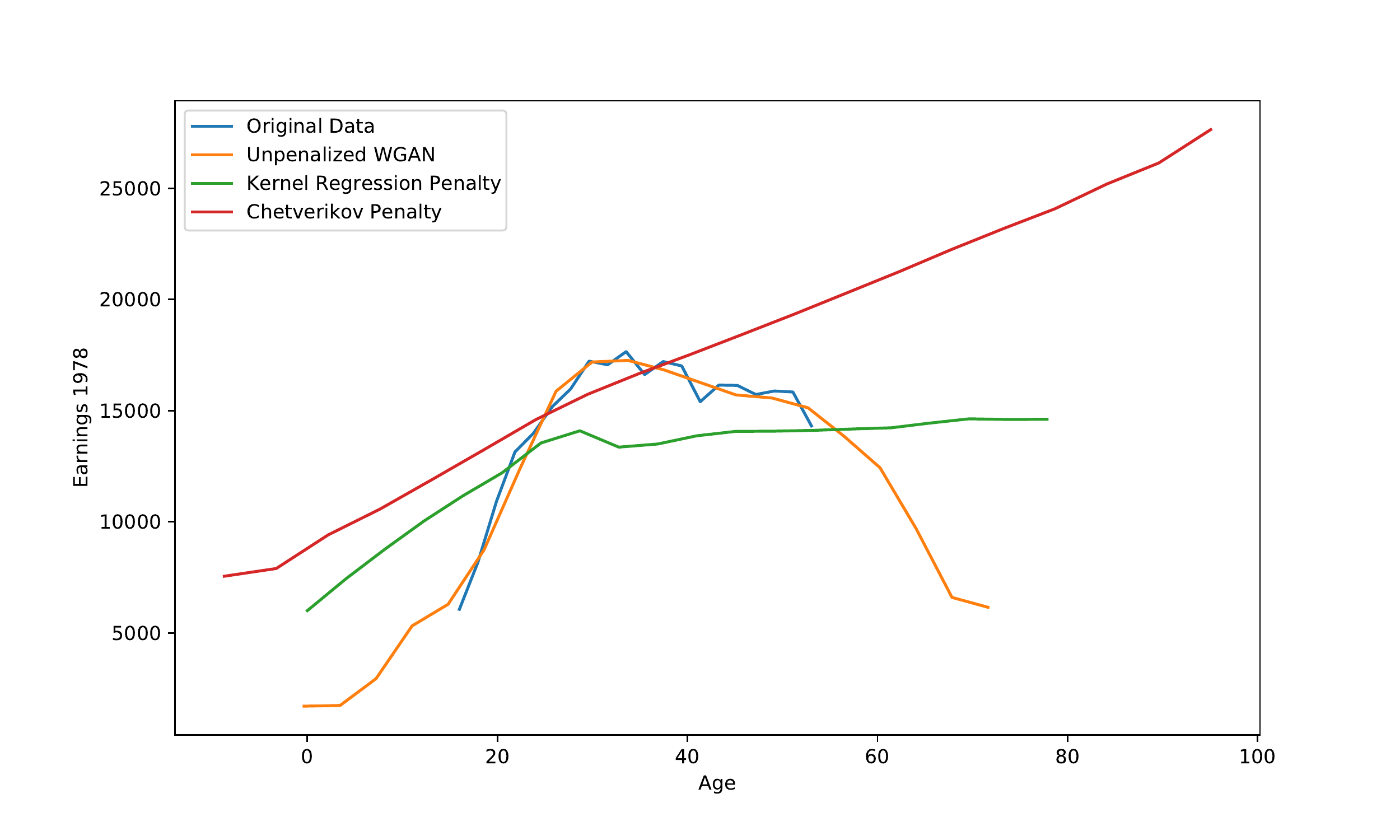}
\caption{Kernel regression of original data and WGANs with various penalties. The regression for the generated data uses more samples, so is smoother.}
\label{fig:monotonicity}
\end{figure}

In addition to the test statistic proposed by Chetverikov, we examine a second penalty. Here, we estimate the conditional mean of interest via a Kernel regression and take the first differences of income along a grid over age, summing all differences which are positive. Although not essential, we only used the gradients of the penalties with respect to earnings '78, i.e. the $Y_i$-variable, when computing the gradients of the loss with respect to the generator parameters via the chain rule during training. We found this to be more stable numerically than using the full gradients during backpropagation, i.e. including the terms involving the derivatives of the penalties with respect to $X_i$, the latter showing up in denominators. The resulting conditional means are shown in Figure \ref{fig:monotonicity}.

\begin{figure}[h!]
\includegraphics[trim=100 50 100 20,clip,width=\textwidth]{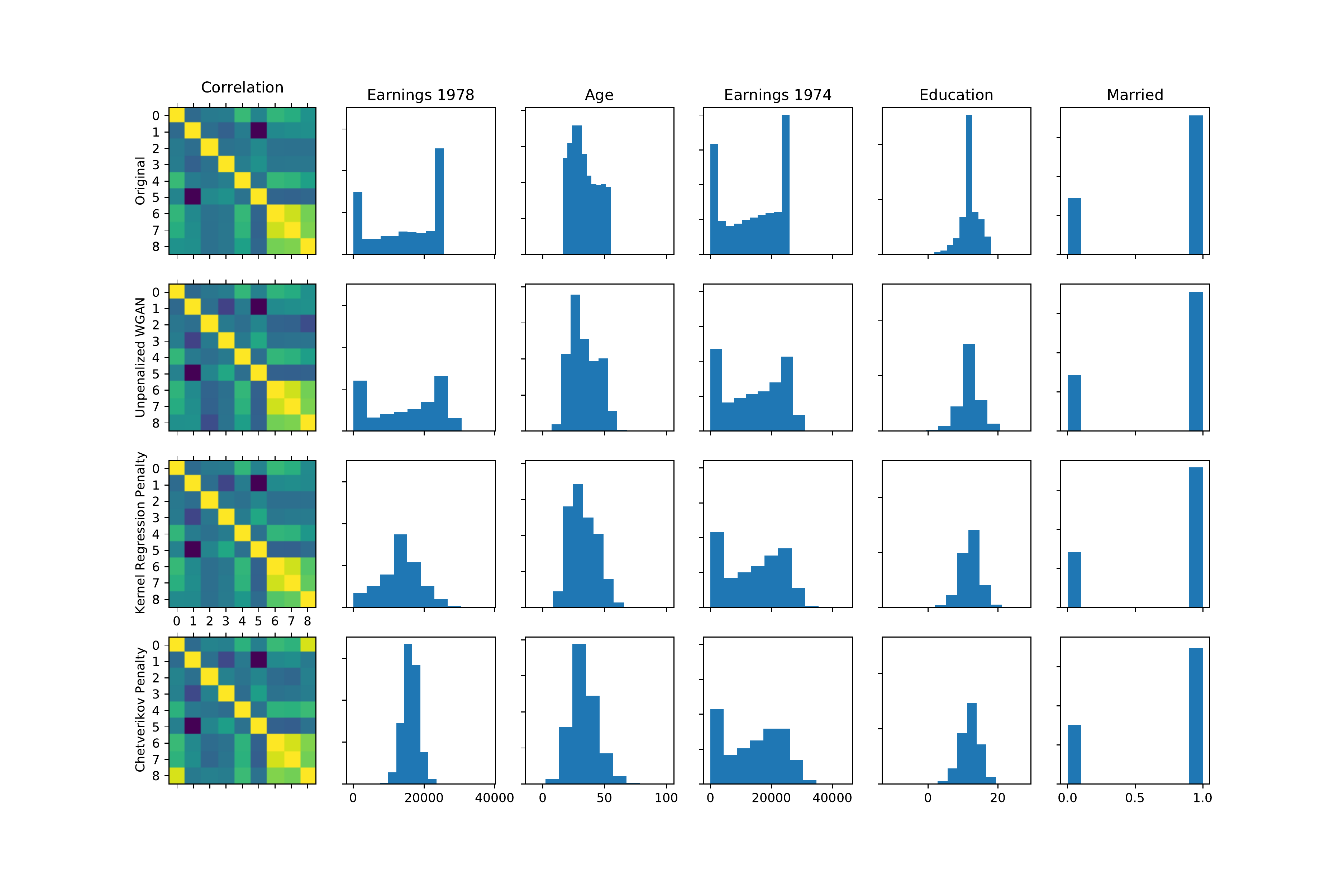}
\caption{Penalized WGANs successfully fit unpenalized aspects of the data}
\label{fig:monofit}
\end{figure}

Both penalties successfully enforce monotonicity despite it being violated in the original data. As shown in Figure \ref{fig:monofit}, we also see that the model learns to capture the non-penalized aspects of the data well. This is particularly surprising, as one might expect to observe a ``vanishing gradients'' phenomenon: the critic might focus mostly on the imposed deviations from the data, which would prevent it from providing meaningful gradients to the generator for the remaining features of the data. This however does not happen, which can be attributed to the stability advantages of the Wasserstein distance over the JS divergence discussed in Section 2. 

However, we see that the Chetverikov test has a much bigger impact on the generated data than the Kernel regression penalty. In particular, it appears to be incentivizing linearity beyond monotonicity. As we see in this example, the properties which are desirable for a test statistic $T(X)$ may or may not be desirable for a penalty $P(X)$ and vice versa. What both have in common is that they should take on large values under violations of $H_1$ and be informative even with a limited number of samples. To see why the latter is important, recall that neural network training benefits from optimization via stochastic gradients, which samples only a few observations from the generator at every iteration. Further, a WGAN penalty should yield informative gradients to the generator, i.e. it should at least be point-wise differentiable wrt. a subset of the generated variables, which is the case for the Chetverikov penalty. Another desirable property for a penalty is that its gradient is zero if $H_0$ is not violated, which the Chetverikov penalty does not satisfy. This is likely the reason of its observed side-effects. A simple fix could be setting $P(X) = max(T(X), C)$, where $C$ is some critical value below we cut off the test statistic. We were not able to improve beyond $C=0$ in our experiments, although it outperformed $C=-\infty$. This is likely driven by the fact that the critical values of the Chetverikov test change with the distribution of the data and thus throughout training, which makes it difficult to limit the penalty's impact via a fixed cutoff while ensuring it is enforcing $H_0$. This suggests that pivotal test statistics should be used instead, if available. Since there is no pivotal test statistic for monotonicity, we showed that one can alternatively obtain a penalty, which is also an implicit test statistic, if a differentiable estimator of the feature of the data that is to be restricted is available, by adding its absolute deviation from a desired value. 
The key takeaway of this exercise is that researchers can directly take any existing test statistic in the econometric literature for a given $H_0$ and plug it into the WGAN objective to obtain a DGP which satisfies $H_0$. Surprisingly, the resulting WGAN algorithm remains stable even for relatively complex, adaptive test statistics, and fits unpenalized aspects of the data well. While existing test statistics can be recommended as good starting points, the researcher may find they affect the model beyond $H_0$, which may require modification or alternative solutions.

\section{Conclusion}

In this paper we show how WGANs can be used to tightly link Monte Carlo studies to real data. This has the benefit of ensuring that simulation studies are grounded in realistic settings, removing the suspicion that they were chosen partly to support the  preferred methods. In this way the simulations studies will be more credible to the readers. We illustrate these methods comparing different estimators for average treatment effects using the Lalonde-Dehejia-Wahba data. 
There are a number of findings. First, in the three different settings, the experimental data, the CPS control group and the PSID control group, different estimators emerge at the top. Within a particular sample the results appear to be relatively robust to changes in the analysis (e.g., changing the sample size, or doing the cross-fitting WGAN robustness analysis). Second, the  preference in the theoretical literature for double robust estimators is broadly mirrored in our results. Although the flexible double robust estimators (using random forests or neural nets) do not always outperform the other estimators, the loss in terms of root-mean-squared-error is always modest, where other estimators often perform particularly poorly in some settings. If one were to look for a single estimator in all settings, our recommendation would therefore be the double robust estimator using random forests or neural nets. However, one may do better in a specific setting by using the WGANs to assess the relative performance of a wider range of estimators. Finally, we showed that WGANs can also be applied to settings in which the researcher wishes to impose restrictions on the implied distribution. This clarifies that, even in settings in which researchers require some control over their simulations, WGANs offer a way to increase the credibility of their results by tying the remaining aspects of their data generating process to real data.

\newpage

\appendix

\centerline{\sc Appendix: The Estimators}

\begin{enumerate}

\item {\sc Difference in Means (DIFF) }

\[\tau^\dm=\frac{1}{N_1}\sum_{i:W_i=1}  Y_i-\frac{1}{N_0}\sum_{i:W_i=0}  Y_i.\]

\item{\sc The  Bias-Adjusted Matching estimator (BCM) }
\begin{enumerate} 
\item Match all treated units with replacement to control units using diagonal version of Mahalanobis matching. 
\item Regress difference between treated and control outcome for matched pairs on difference in covariates. See \citet{abadie2006, abadie2011bias}.
\end{enumerate} 

\item{\sc Conditional Outcome Model, Linear Model (LIN)}: See $\hat\tau^{\cm,\lm}$ in \ref{subsection:definitions}.

\item{\sc Conditional Outcome Model, Random Forest (RF)}: See $\hat\tau^{\cm,\rf}$ in  \ref{subsection:definitions}. 

\item{\sc Conditional Outcome Model, Neural Nets (NN)}: See $\hat\tau^{\cm,\nnn}$ in \ref{subsection:definitions}.

\item{\sc The Horowitz-Thompson Estimator, Logit Model (LIN)}: See $\hat\tau^{\hoth,\lm}$ in \ref{subsection:definitions} and \citet{hirano2003efficient}.

\item{\sc The Horowitz-Thompson Estimator, Random Forest (RF)}: See $\hat\tau^{\hoth,\rf}$ in \ref{subsection:definitions}.

\item{\sc The Horowitz-Thompson Estimator, Neural Net (NN)}: See $\hat\tau^{\hoth,\nnn}$ in \ref{subsection:definitions}.

\item{\sc The Double Robust Estimator, Linear and Logit Model (LIN)}: See $\hat\tau^{\dr,\lm}$ in \ref{subsection:definitions}.

\item{\sc The Double Robust Estimator, Random Forest (RF)}: See $\hat\tau^{\dr,\rf}$ in \ref{subsection:definitions}.

\item{\sc The Double Robust Estimator, Neural Nets (NN)}: See $\hat\tau^{\hoth,\nnn}$ in \ref{subsection:definitions}.

\item{\sc Residual Balancing Estimator (RB)} 
\begin{enumerate} 
\item Estimate conditional outcome mean for controls by elastic net. \item Construct weights that balance control covariates to average covariate values for treated.\item Combine to estimate average outcome for treated units under control treatment. See \citet{athey2018approximate}.

\end{enumerate} 

\item{\sc Causal Forest Estimator (CF)} 
See \cite{athey2019generalized}. 
\end{enumerate}

\appendix

\centerline{\sc Appendix: The Adam Algorithm}

\begin{algorithm}[htbp]
\caption{Adam}\label{algo:adam}
\begin{algorithmic}[1]
\\ {$\rhd$ Tuning parameters: }
\\ \hskip0.6cm {$m=$, batch size}
\\ \hskip0.6cm {$\alpha$, step size}
\\ \hskip0.6cm {$\beta_1$, }
\\ \hskip0.6cm {$\beta_2$, }
\\ \hskip0.6cm {$\epsilon=10^{-8}$, }
\\ {$\rhd$ Starting Values: }
\\ \hskip0.6cm {$\theta=0$, $m_0=0$, $v_0=0$, $t=0$}
\While{$\theta$ has not converged}
\State $\rhd$ $t\gets t+1$
\State Sample $\{Z_{i}\}^{m}_{i=1}$.
\\
\hskip0.6cm  $\rhd$ {Compute gradient}
\State $\delta_{\theta} \gets\frac{1}{m}\sum_{i=1}^m \nabla_{\theta} f\left(
Z_{i};\theta_t\right)$
\State $\gamma_\theta\gets\frac{1}{m}\sum_{i=1}^m\left( \nabla_{\theta} f\left(
Z_{i};\theta_t\right)\right)^2$
\State $m_t\gets \beta_1 m_{t-1}+(1-\beta_1) \delta_\theta$
\State $\hat m_t=m_t/(1-\beta_1^t)$
\State $v_t\gets \beta_2 v_{t-1}+(1-\beta_2)\gamma_\theta$
\State $\hat v_t=v_t/(1-\beta_2^t)$
\hskip0.6cm $\rhd$ {Update $\theta$}
\State $ \theta_t\gets\theta_{t-1}-\alpha\hat m_t/(\sqrt{\hat v_t}+\epsilon)$
 (update generator parameter)
 \State {\bf end while}
\EndWhile
\end{algorithmic}
\end{algorithm}

\newpage

\bibliographystyle{plainnat}
\bibliography{references}

\end{document}